# GEM-based detectors for direct detection of low-mass WIMP, solar axions and narrow resonances (quarks).

B.M.Ovchinnikov, V.V.Parusov*

Institute of Nuclear Research of Russian Academy of Sciences, Moscow.

*Corresponding author; e-mail: parusov@inr.ru

**Abstract**

Gas electron multipliers (GEMs) with wire (WGEMs) or metal electrodes (MGEMs), which don't use any plastic insulators between electrodes are created (see sections 2 - 4). The chambers containing MGEMs (WGEMs) with pin-anodes are proposed as detectors for searching of spin-dependent interactions between Dark Matter particles and gases with nonzero-spin nuclei ($^{1}H$, $D_2$, $^{3}He$, $^{19}F$, $^{21}Ne$, $^{129}Xe$, $^{131}Xe$, etc.). In this paper, we present a review of such chambers. As photosensitive addition we use of $H_2$, $CH_4$, $CF_4$, $C_2H_4$,TMAE, etc.

In our experiments as another filling of the chambers for search of low-mas WIMP (<10 GeV/$c^2$), solar neutrino and solar axions with spin-dependent interaction we propose to use the mixtures: $D_2$ + 3ppmTMAE, $^{3}He$ + 3%$CH_4$, $^{21}Ne$ + 10%$H_2$, at pressure 10-17 bar. And in our experiment with liquid mixtures is used the mixtures with $^{19}F$ (LXe + $CF_4$) and mixture with $^{1}H$ ($LCH_4$ + 40ppm TMAE).

For investigation of the gas mixtures Ne+10%$H_2$, $H_2$ ($D_2$) +3ppmTMAE, the chamber containing WGEM with pin-anodes detection system was constructed (section 5). In this paper we present the results of an experimental study of these gaseous mixtures exited by an α - source. Mixture of Ar + 40 ppm $C_2H_4$ and mixture 50% Xe + 50%$CF_4$ have been investigated. The spatial distributions of photoelectron clouds produced by primary scintillations on α- and β-particle tracks, as well as the distributions of photoelectron clouds due to photons from avalanches at the pin-anode, have been measured for the first time (section 6).

In our experiment with liquid mixtures (section 7) is used the mixtures with $^{19}F$ (LXe + $CF_4$) and mixture with $^{1}H$ ($LCH_4$ + 40ppm TMAE).

The time projection chamber (TPC) with the mixture $D_2$ + 3ppmTMAE filling at a pressure 10 bar allow to search of spin-dependent interactions of solar axions and deuterium (section 8). As well as we present the detecting systems for search of narrow pp-resonances (quarks) in accelerators experiments (section 9).

In our experiments the electronegative impurities ($O_2$, $C_2F_4$, $C_3F_8$ etc.) were removed from the gases mixtures an a multistage purification system to a level of $10^{-8}$ $O_2$ equivalent (0,01 ppm). The entire system (chamber+ gas system) was checked by the "ISTOK" gas analyzer for the presence of known electronegative impurities.

## 1. Introduction

More than forty years ago G. Charpak and F. Sauli have introduced their Multi-Step Chambers to overcome limitations of gain in Parallel-Plate and Multi-Wire Proportional Chambers (MWPC) [1, 2]. These MWPCs have revolutionized detection systems in high energy physics.

Currently there are different types of detectors for fast detection and localization of charged particles exist. One of them is a Gas Electron Multiplier (GEM). A standard Gas Electron Multiplier [2,3,4] consists of a thin composite sheet (plate) with two metal layers separated by a thin insulator and pierced by a regular matrix of open channels. These plates contain through holes on all their area, separation distances and diameters of which are approximately equal to the plate thickness (Fig.1a). Inside these holes, which are filled with corresponding gases, in presence of strong electric fields, a multiplication of electrons takes place. GEMs provide the best spatial resolution and higher rate than the wire chambers (MWPC). More coarse macro-patterned detectors are thick-GEMs (THGEM) [5, 6, 7] or patterned resistive thick GEM devices (RETGEM) [8].

However, the most essential disadvantage of GEMs consists in their low reliability and stability. The matter is that in a process of dispersion of the GEM's cathode electrodes by positive ions of proportional avalanches in GEM with metal or high-resistive electrodes (RETGEM). The sedimentation of the sprayed carrying-out material on the walls of holes with subsequent leaks and breakdowns between electrodes takes place. It leads to subsequent decrease of the potential difference between GEM's electrodes and corresponding reduction of the multiplication factor in an avalanche (Fig.1b).

Micro-pattern gaseous detectors (MPGD), due to their tiny electrode structure and small avalanche gaps (Fig.1c), are very fragile and can be easily damaged by sparks appearing at high operational gains (typically at gains of $10^4$ or slightly more) [7].

Therefore, we were concentrated on development of more robust designs of GEM detectors with wire (WGEM) or metal electrodes (MGEM). The idea of WGEM without plastic insulators was first mentioned in our work [9-12]. In our subsequent works [13] a MGEM with metal electrodes of diameter of 22 mm was designed and tested. In the paper [14] we have described a novel concept of MGEMs. In our next works [15-18, 21, 25] it was suggested that the search for spin-dependent WIMP-nucleon interactions with help of detecting system GEM + pin-anodes can be performed (sections 5-7).

To suppress the β, γ and n0 backgrounds for direct detection of low-mass WIMP, we proposed a addition in gases of photosensitive dopants and a comparison of scintillation (S1) and ionization signals (S2) for every event is suggested. The results obtained in our study suggest that it is possible to develop large volume detectors capable of detecting scintillations with a 100% geometrical efficiency, by contrast to the well known detection techniques based on photomultipliers having efficiency of only a few percent [18](section 5.3).

The particles of WIMPs with large masses (> 10 GeV) experimentally were not found so far [24, 27, 29, 31], it is necessary to search the WIMP with small masses (≤ 10 GeV/$c^2$). The LUX experiment finally ruled out the presence of a WMP mass greater than 10 GeV/$c^2$ [52].

a) 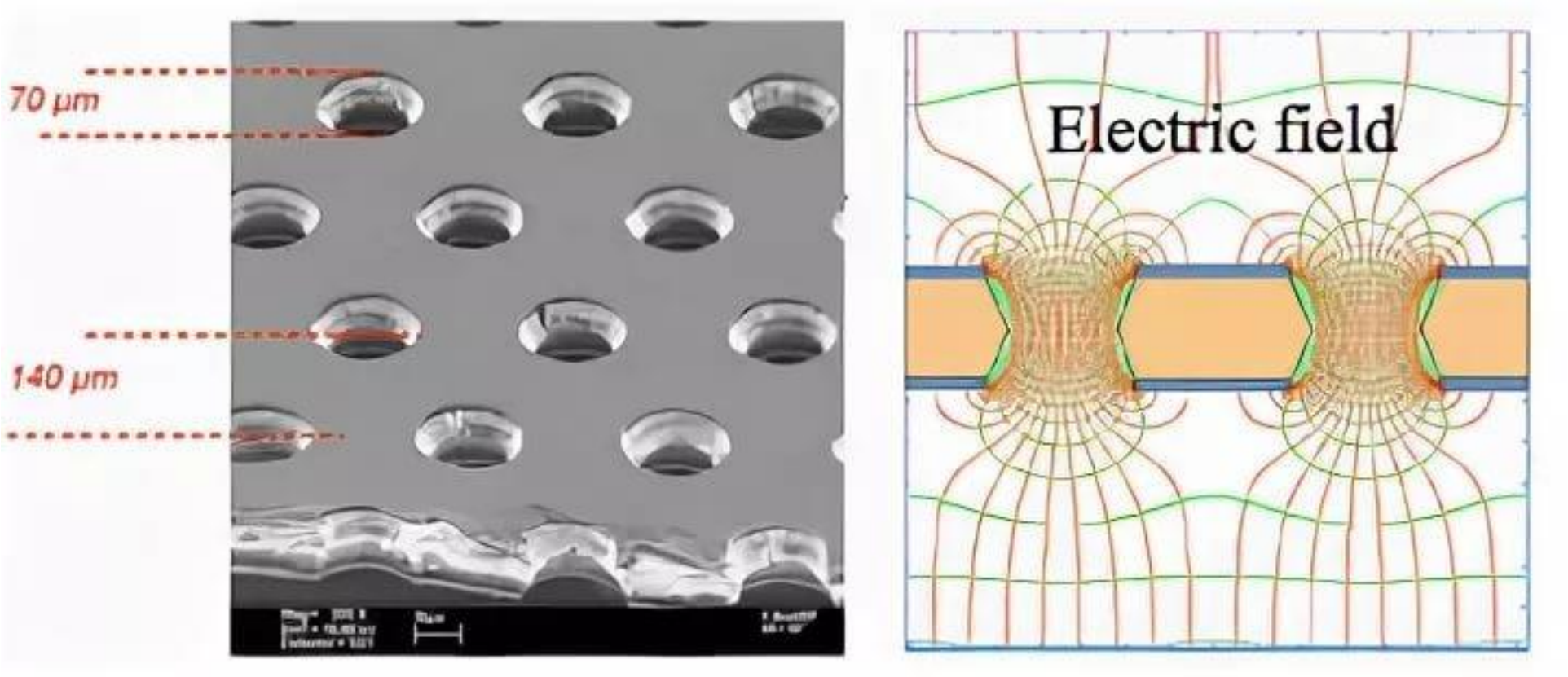

b) 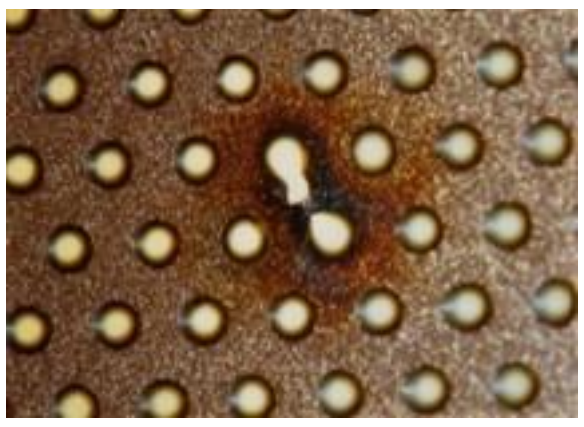

c) 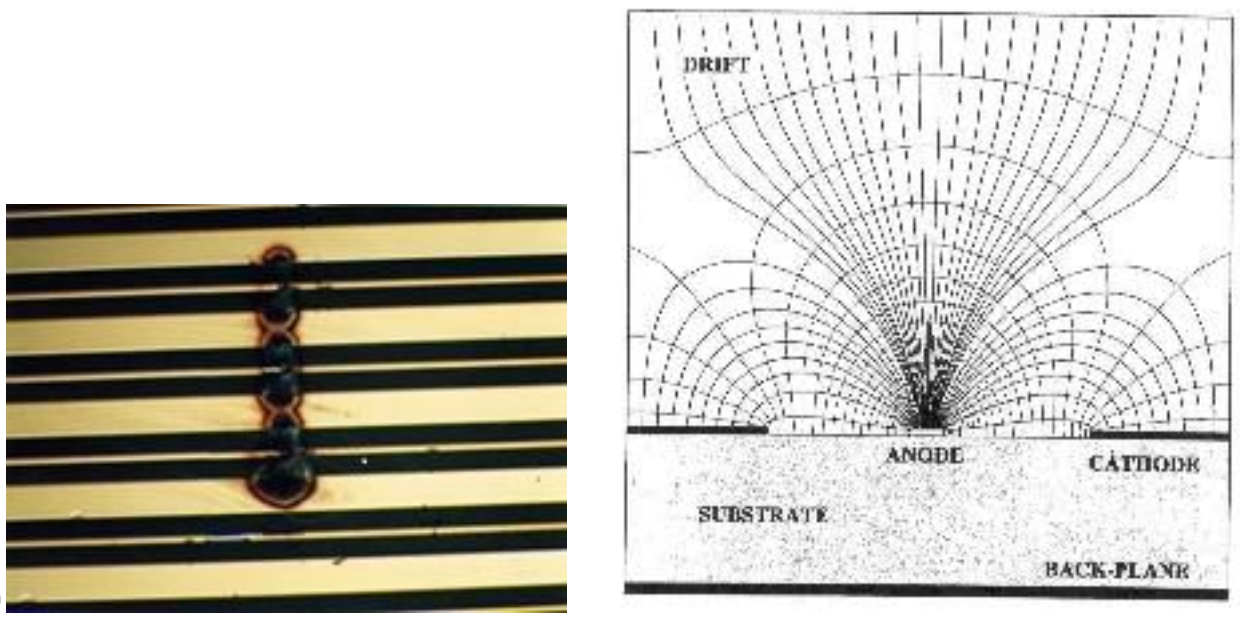

Fig.1**. a)** Gas Electrons Multiplier (GEM), **b)** Resistive thick GEM (RETGEM**), c)** Micro-pattern gaseous detectors (MPGD). The sedimentation of the sprayed carrying-out material on the walls of holes (insulators) with subsequent leaks and breakdowns between electrodes takes place (b, c).

## 2. GEMs with wire electrodes (WGEM)

In our works [9-12] WGEMs with wire electrodes and no plastic insulators between them were created. The WGEMs [9] used macroscopic windows of size 1 mm by 1 mm, while WGEMs [10-12] used windows of size 0,5 mm by 0,5 mm. The gap between the wire electrodes was equal to 1 mm. The design of wire GEMs and the results of their tests are shown in Fig.2 and Fig.3.

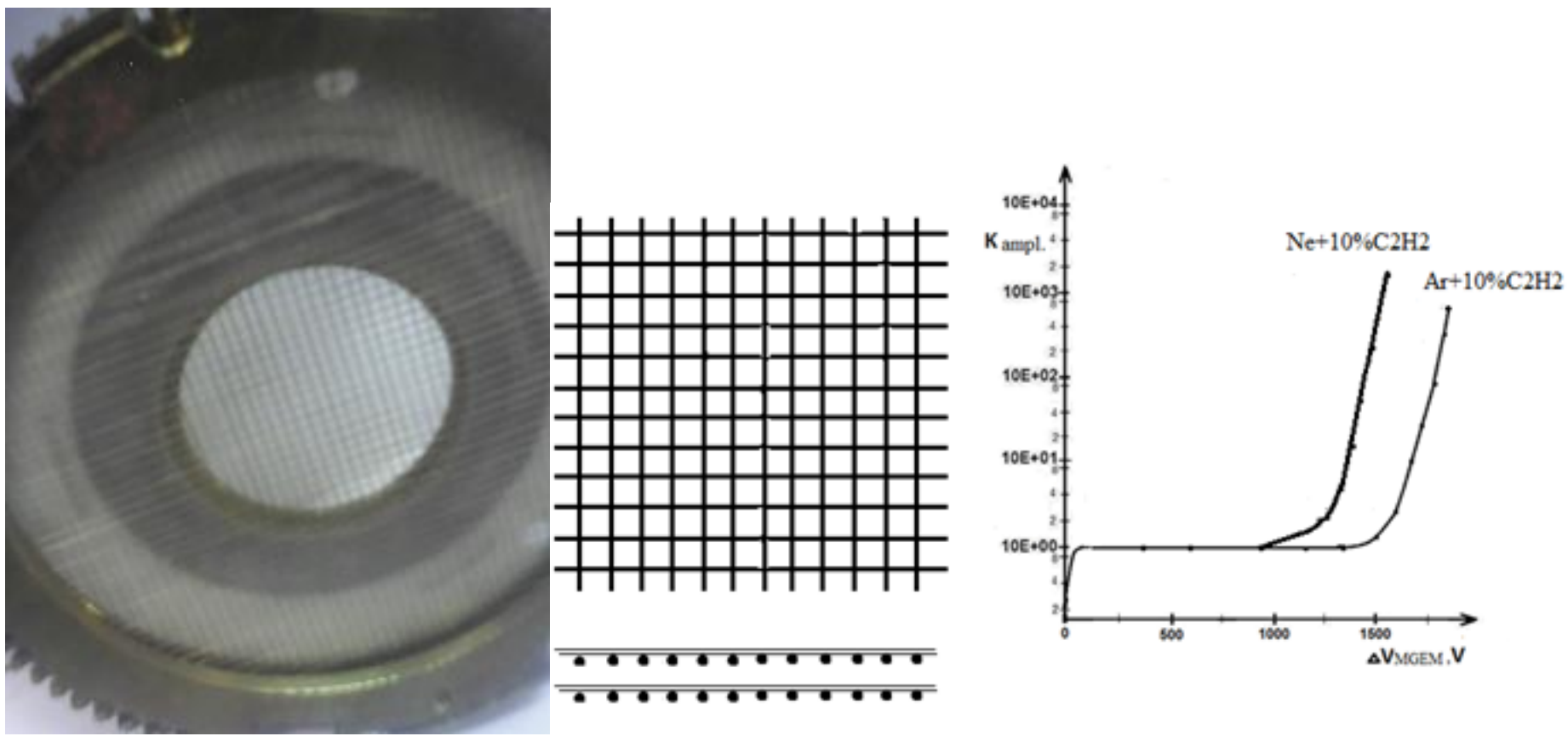

Fig.2. Design of WGEMs and results of their tests [9].

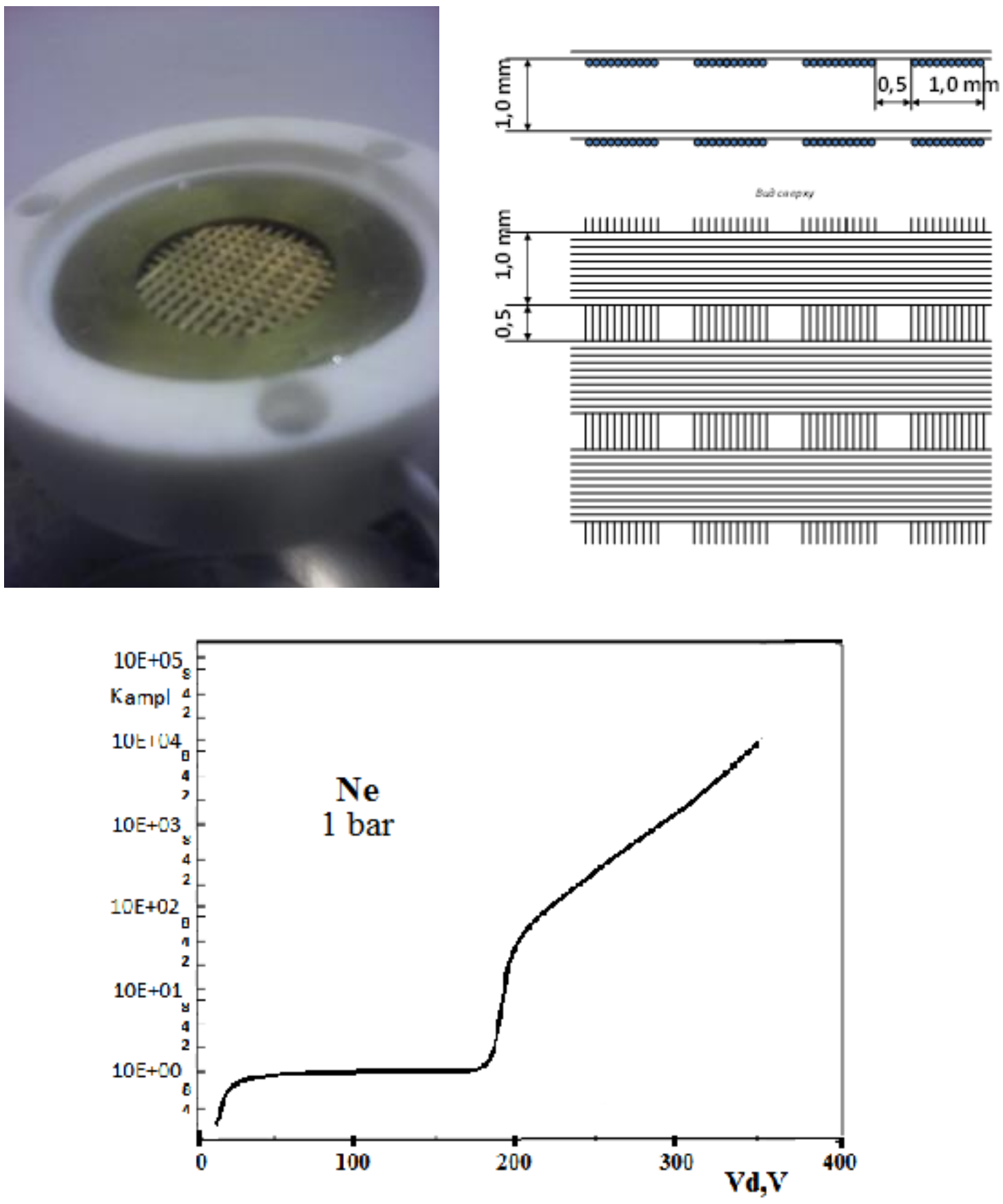

Fig.3. Design of WGEMs and results of their tests [10-12].

### 3. GEM with metal electrodes (MGEM)

In the work [12] a MGEM with metal electrodes and sensitive area of 22 mm by 22 mm for the first time was demonstrated. The electrodes were made by drilling of 1 mm holes with a step between them of 1.5 mm in to 1 mm thick brass plates (see Fig.4). One disadvantage of that MGEM [13] is large duration of the process of drilling the holes in the electrodes, especially for a case of small diameters and small steps between holes. In addition to that, formation of agnails at the edges of holes in a process of drilling is possible. Another problem of drilling of holes with drilling machines consists in difficulty of production of large area MGEM electrodes with high accuracy of sizes of holes and their positions on various plates.

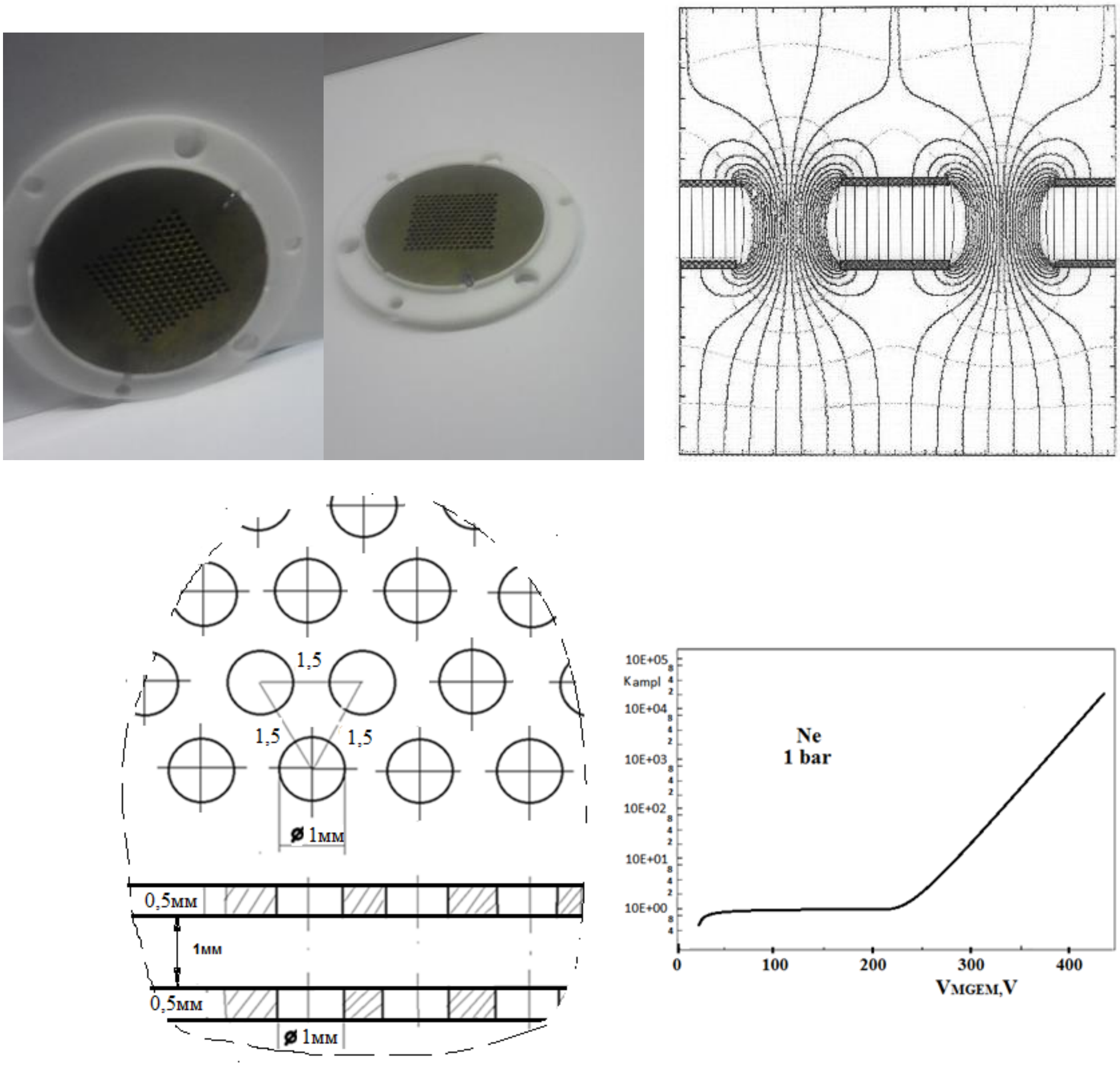

Fig.4. Design of metal GEM, electric field and results of tests [13].

## 4. MGEM with the etching of holes

In the work [14] the gas electron multiplier with metal electrodes (MGEM), differing by its simplicity, high precision and technological effectiveness in production, as well as reliability and stability of its operation, is made and tested. To eliminate certain specific shortcomings (see Introduction), in this work metal electrodes were made by a method of drawing a mask on 0.3 mm thick brass plates with 0.3 mm diameter holes and 0.5 mm step between them (fig.5) with subsequent double-side etching of the holes.

As far as the initial brass plate was cut from 0.3 mm thick rolled foil, the produced electrodes had a curved shape. Therefore, at assembling, between the MGEM electrodes a fluoroplastic spacer was introduced to increase resistance on the pass of leakage charges between the electrodes. From the outer sides both electrodes were pressed to fluoroplastic plate by additional steel rings of 2 mm thickness.

By means of central 4 mm holes in electrodes and special insulating bolts, the GEM electrodes were mutually positioned in such a way that the relative displacements of holes didn't exceed 0.02 mm. The gap between electrodes was chosen to be equal to 1 mm. In this design a sensitive area GEM with holes had a diameter of D=75 mm. The GEM was tested in the chamber (Fig.6.) with various filling gases: Ar +10%$C_2H_4$ (1 and 0.4 bar), Ne +(($O_2$ +$N_2$ +$H_2O$)·$10^{-6}$)(1 and 0.4 bar), Ar +10% $CH_4$ (1 bar).

The results of tests at irradiation of the drift gap of the chamber by alpha-particles ($Pu^{239}$ ) are presented in Fig.5. It is visible, that Ne provides the maximum coefficient of multiplication at the smallest potential difference between electrodes before the breakdown happens.

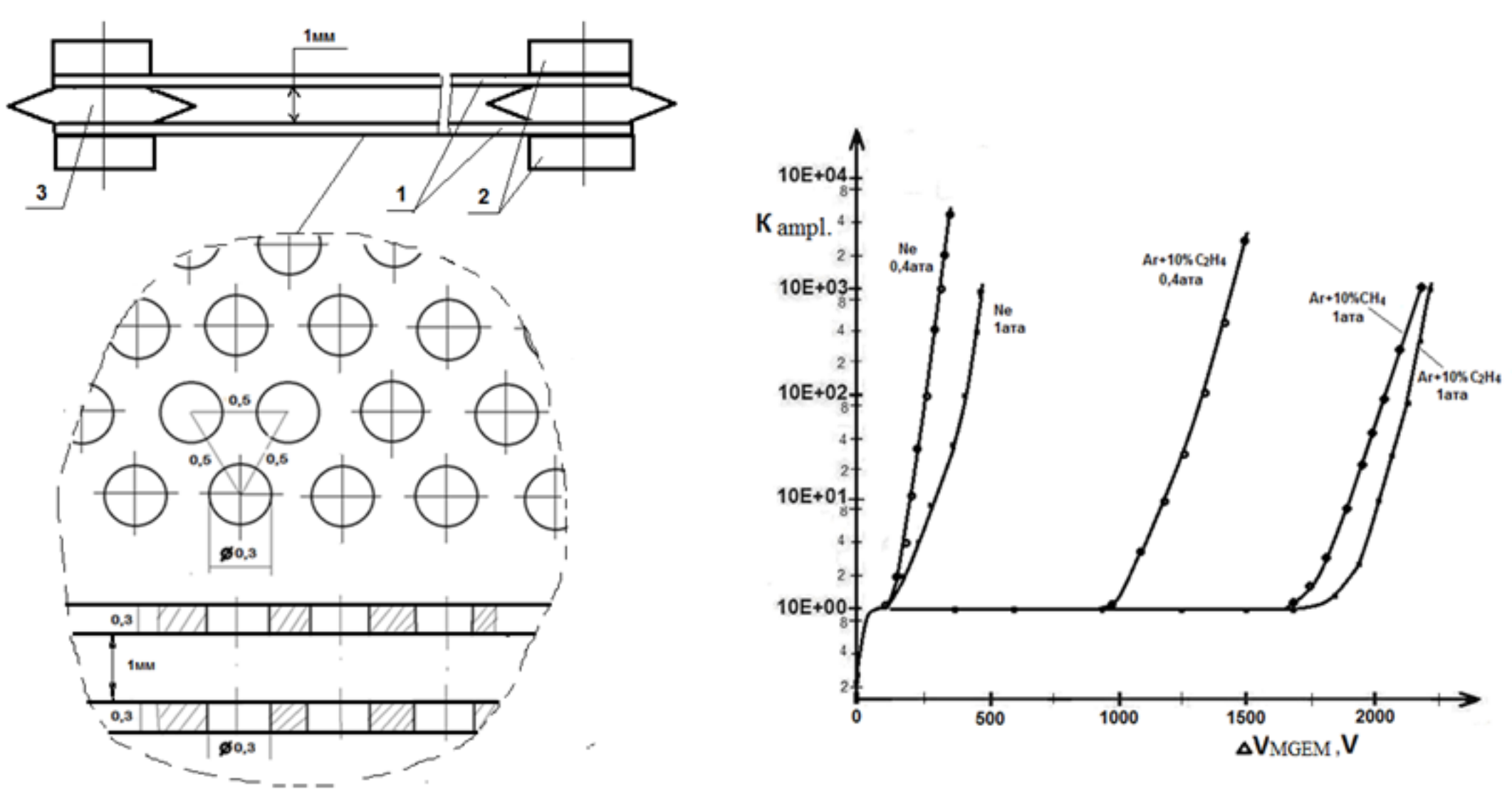

Fig.5. Left: design of MGEM: 1-plates of MGEM, 2-clamping rings, 3-layer from fluoroplastic. Right: amplification of MGEM filled with different gases.

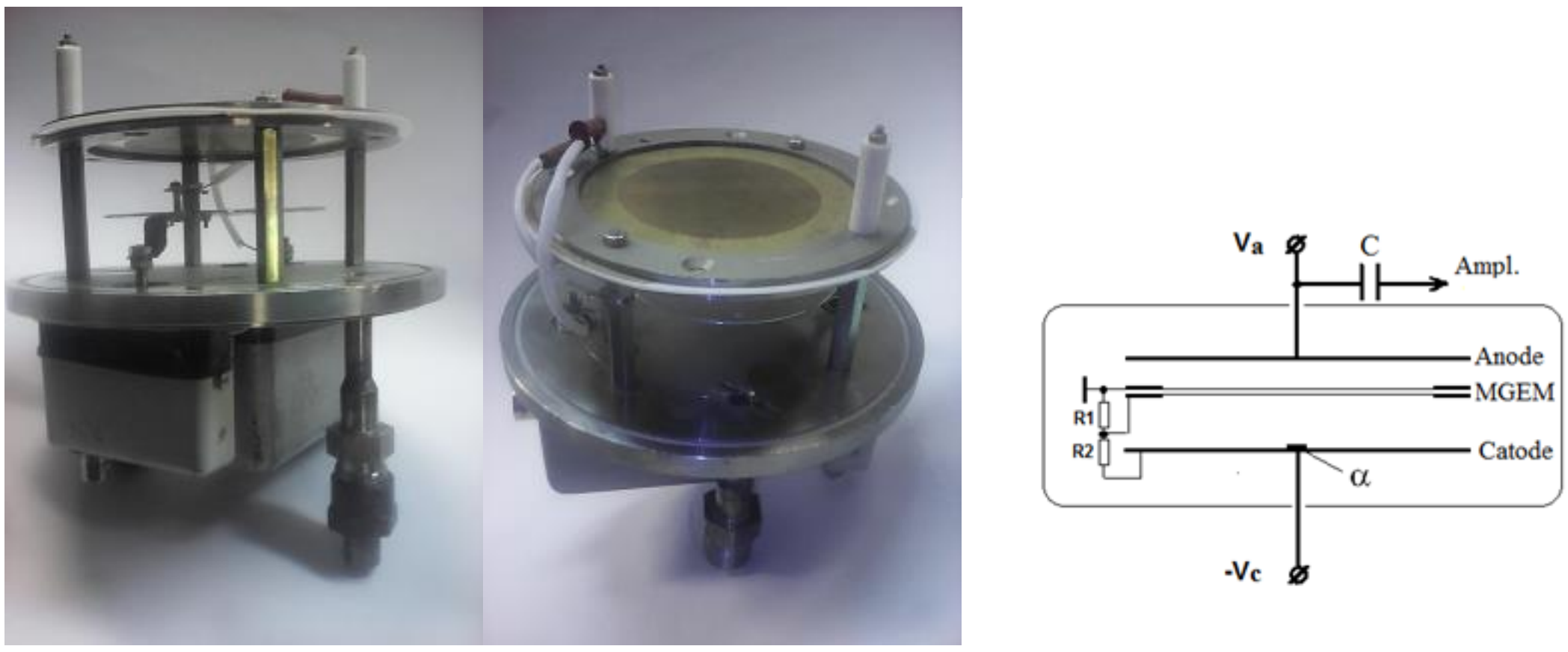

Fig.6. Chamber for test MGEM.

**5. The detecting chambers with system GEM + pin-anodes for direct detection of WIMP**.

In work [15] the idea of focusing screen with holes + system pin-anodes already was shown (section 5.1). The wire gas electron multipliers in combination with pin-anodes are proposed for detection of events:

(1) In the gas phase of a double-phase argon chamber (section 5.2);
(2) In chamber for direct detection of WIMP with mass ≤ 0.5 GeV/$c^2$ (section 5.4);
(3) In chamber with gas mixture $H_2$**(**$D_2$) + 3ppm TMAE for direct detection of WIMP with mass ⩽ 10 GeV/$c^2$ and solar axions (section 5.5);
(4) In double-phase chambers with LXe + $CF_4$ and $LCH_4$ + TMAE filling (section 7).
(5) In scintillation and ionization fast chambers with mixture Xe + $CF_4$ for accelerators experiments (section 9).

In the time projection chamber (TPC) with the mixture $D_2$ + 3ppmTMAE filling we use of two-step system GEM + MWPC (section 8).

**5.1 A liquid-methane ionization chamber**

A liquid-methane ionization chamber with a system of focusing screen + pin-anodes is proposed as a setup to search for spin-dependent interactions of DM particles [15]. The anode of the chamber is placed in gaseous methane above liquid methane. The anode consists of a system of pins. The Focusing screen is placed between the Anode and liquid methane. The screen has a system of holes concentrically on the relevant pin-anode. The values of electrical potentials on the electrodes of the chamber are set in such a way that all electrical lines of force are focused on the pin-anodes. We provide only the idea of using a specially designed liquid-methane ionization chamber in an experiment aimed at searching for the (mostly) low mass DM based on their spin-dependent interactions (see section 7.2).

### 5.2 Double-phase argon chamber

Multichannel WGEM + system pin-anodes are proposed for detection of events in the gas phase of a double-phase argon chamber [16, 17]. Hydrogen with a concentration of 10 % is added to argon to eliminate feedbacks via photons emitted by excited argon molecules in avalanche development processes during detection of events in the gaseous argon. A maximum electron multiplication coefficient of ∼300 has been obtained for the multichannel wire gas electron multipliers with a 1 mm gap used to detect α-particles in the Ar + 10% $H_2$ mixture at a pressure of 1 bar. When a pin anode is used, the maximum electron multiplication factor for α-particles is $\sim 2.5 \times 10^5$. It has been experimentally shown that adding $H_2$ with a concentration of 100 ppm to liquid argon (10%$H_2$ in gas phase) has no effect on the singlet component of the scintillation signal in the liquid argon and reduces the emission efficiency relative to the pure argon gas phase only slightly (by 20%).

### 5.3. A method for background reduction in experiments for direct detection of WIMPs. Investigation of a mixture of Ar + 40ppm $C_2H_4$.

To suppress the β, γ and n0 backgrounds, we proposed [17, 18] a addition in liquid argon of photosensitive dopants and a comparison of scintillation (S1) and ionization signals (S2) for every event is suggested. The addition in liquid Ar of photosensitive TMA, TMG or $C_2H_4$ [19] and suppression of triplet component of scintillation signals ensures the detection of scintillation signals with high efficiency and provides a complete suppression of the electron background.

In work [20] we investigated of scintillation (S1) and ionization signals (S2) on a mixture of Ar + 40ppm $C_2H_4$ at a pressure of 5 bar.

The measurements were taken inside the chamber similar a chamber with pin-anode (see Fig. 5 left). The mixture was irradiated with α (239Pu) and β (63Ni) particles. The chamber was used with potentials of pin-anode Va = 1300 V, Kamp ~$10^4$ (β) and Va = 520V, Kamp = 30 (α).

Peak S1 is associated with the cloud of photoelectrons from the chamber volume due to scintillation photons with λ = 128 nm, which are emitted upon excitation of argon atoms with α(β) particles: $h\nu + C_2H_4 \rightarrow C_2H_4^* + e^-$. Peak S2 is due to ionization electrons from α(β) particles (Fig. 5a). Peak S2 fluctuates in the amplitude for different events, since the β-particle spectrum lies in the range of $E^{min}$ - $E^{max}$ (17 - 67 KeV). Peak S3 can be attributed to the cloud of photoelectrons from the chamber volume produced by photons from avalanches at the pin-anode.

In Fig. 5b we have shown of signal forms for α-particles in the case of using a broadband amplifier with integration and differentiation constants equal to 1 millisecond on a chip KR140UD26a instead of a charge-sensitive amplifier BUS 2-96. The ionization peak S2 position fluctuates with time for different events.

The results obtained in our study suggest that it is possible to develop large volume detectors capable of detecting scintillations with a 100% geometrical efficiency, by contrast to the well known detection techniques based on photomultipliers having efficiency of only a few percent [27].

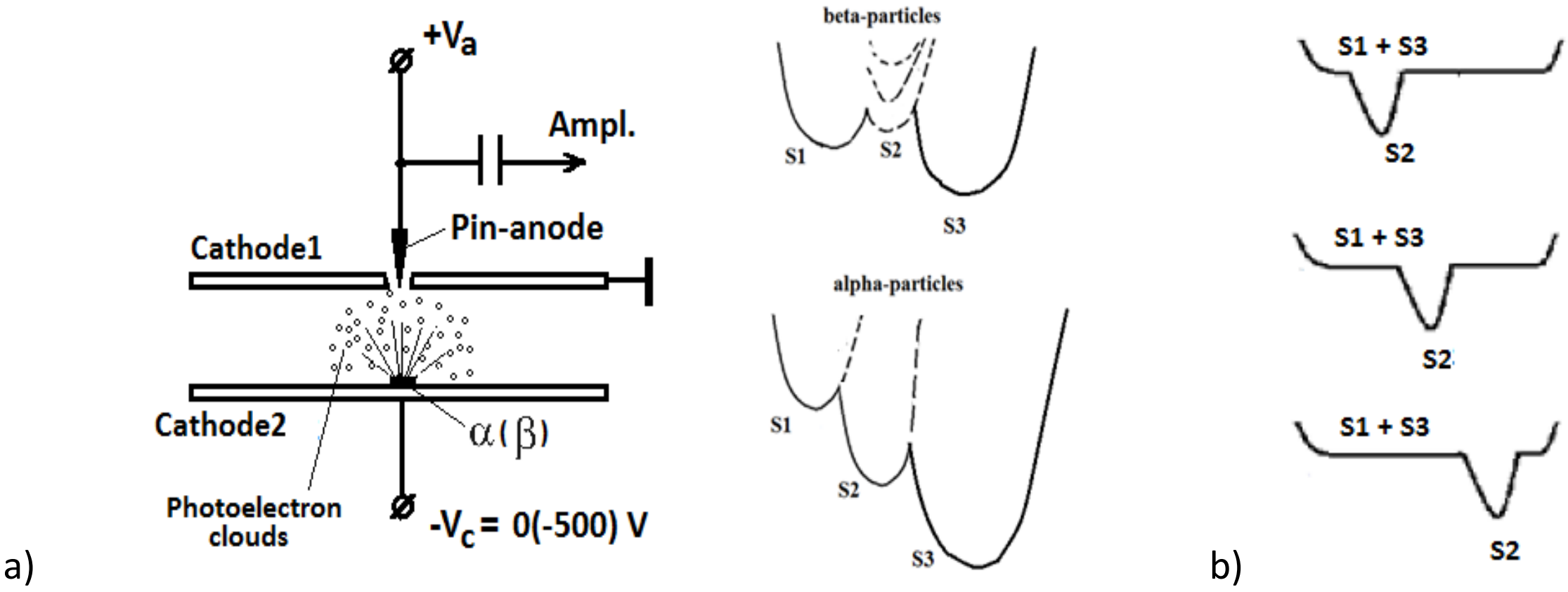

Fig.5. a) Chamber filled with the mixture of Ar + 40ppm $C_2H_4$ at a pressure of 5 bar. The mixture was irradiated with α (239Pu) and β (63Ni) particles. The chamber was used in potentials of pin-anode Va = 1300 V, Kamp ~$10^4$(β-particles) and Va = 520V, Kamp = 30 (α-particles), –Vc = 0V. We used a charge-sensitive amplifier BUS 2-96; b) Signal form for α-particles (239Pu) in the case of using a broadband amplifier on a chip KR140UD26a.

### 5.4 The chamber for direct detection of WIMP with mass ≤ 0.5 GeV/$c^2$.

The detectors with pure NaI, Xe or Ar [24, 31] make it possible to search WIMPs with large masses (up to dozens or hundreds GeV), as far as the energy of nuclear recoils in these detectors from low mass WIMPs is low. To account for yearly modulation effect in DAMA-LIBRA experiment [28, 29] J.Va'vra has supposed [31] that this effect is explained by low mass WIMP scattered at protons in $H_2O$ molecules, which is contained in NaI crystals at about 1ppm level (see Table 1.)

The chamber for direct detection of WIMP with mass ≤ 0.5 GeV/$c^2$ was developed [21]. The chamber (see Fig.7) is filled with gas mixture Ne+10% Hydrogen +0,15ppm TMG. In this chamber for the events detection it was used a system GEM +pin-anodes, which provide the energy threshold about eV. The electron background is suppressed due to photosensitive addition of TMG. For a direct detection of WIMP it is proposed to use a liquid argon chamber with Hydrogen dissolved in liquid argon at a concentration 100ppm+0,015ppm TMG. Based on the work [22], where in a spherical proportional detector the energy threshold is about 100eV, while the amplification factor of the detecting system is about $10^4$, we estimate the threshold of our experiment to be about ~100 eV·$10^4$/5·$10^7$ < 1eV. The $H_2$-filling provides an efficient suppression of the electron background, because of the short track of recoil protons, compared to the one from background electrons [30]. As another filling of the chamber is used the mixture $^{21}$Ne+10% $H_2$ for search of spin-dependent interaction [23, 35, 36].

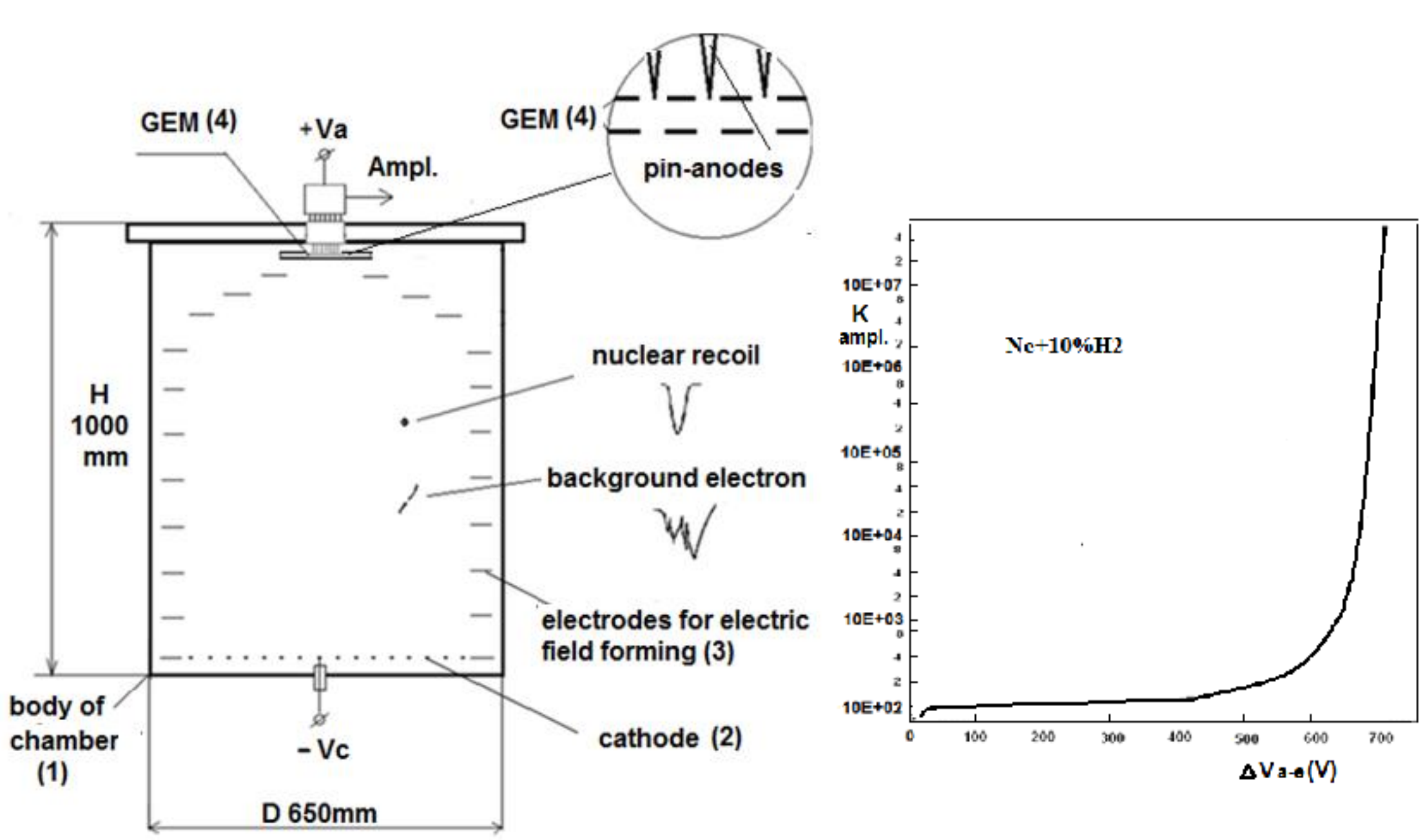

Fig.7. Chamber with a system GEM +pin-anodes and corresponding results of tests of the detection system (Fig.8). Coefficient of electron multiplication as a function of the anode voltage. The chamber is filled with a mixture of Ne+10%$H_2$ at a pressure of 1 bar is exposed to α particles.

**5.5 The study of system GEM + pin-anode for search for low-mass WIMP and solar axions.**

In work [49] we have proposed first time the experiment with deuterium ($D_2$) for direct detection of solar neutrino.

In our work [25] the chamber for direct detection of WIMPs with mass < 10 Gev/$c^2$ and axions, emitted from the Sun, was developed. For searching the solar axions is used of mixture $D_2$ + 3 ppm TMAE in this experiment. The chamber is filled with a gas mixture $H_2$ +3ppm TMAE (5, 10 bar), or $D_2$ + 3ppm TMAE (5, 10 bar). These gas fillings allow to suppress the electron background [18]. For detection of events is used a system GEM + pin-anodes (Fig.8) with coefficient multiplication of about $10^5$ (see Fig.9) and the chamber of the previous experiment (Fig.7). Collisions of WIMPs with $H_2$ provide recoil protons with energies of several keV (see Table 1). An addition of TMAE with a low ionization potential (5,36 eV) provides detection of recoil protons.

As another filling of the chamber is used the mixture deuterium ($D_2$) + 3ppm TMAE, because:

(1) Energy of D-recoil is two times more than proton ($^1$H) recoil.

(2) Spin ($J_S$) of nuclear deuterium is equal 1 ($J_S$ of nuclear hydrogen is equal ½) [23]. This gives the increase the cross section 3-times as compared with hydrogen.

Scattering cross section is $\sigma_{SD} \sim J_S \cdot (J_S + 1)$. This gives the increase the cross section 3-times as compared with hydrogen [15]. The $H_2$-filling provides the electron background suppression, because the recoil protons in $H_2$-medium have the short track [30], as distinguished from background electrons.

Perhaps the energy of axion is equal to several keV [26], is transfers the energy to recoil deuterium. The $H_2$-filling provides the electron background suppression, because the recoil protons in $H_2$-medium have the short track [30], as distinguished from background electrons. The chamber with $D_2$ + 3ppm TMAE filling are placed in low background laboratory in low background shielding for search the yearly or daily modulation effects.

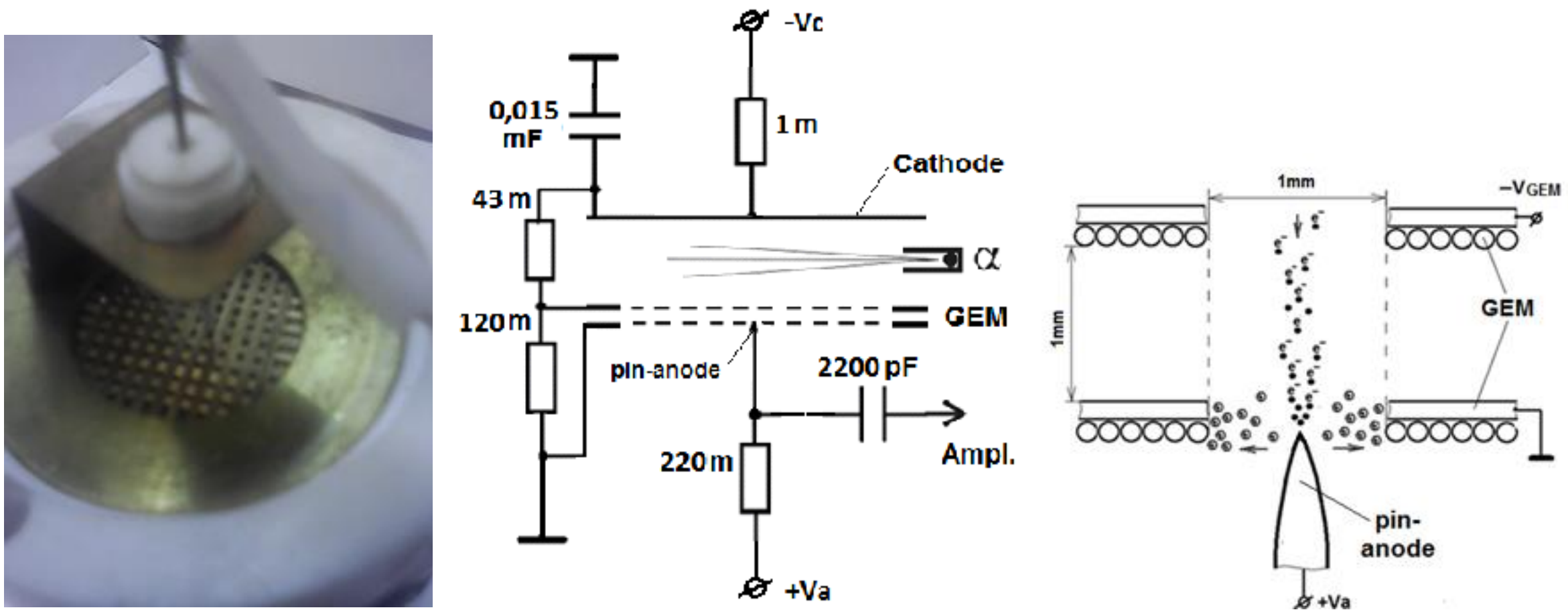

Fig.8. Detection system for testing of WGEM + pin-anode with diagram of the travel of positive ions from avalanches developed at the pin and electrons being collected at the pin.

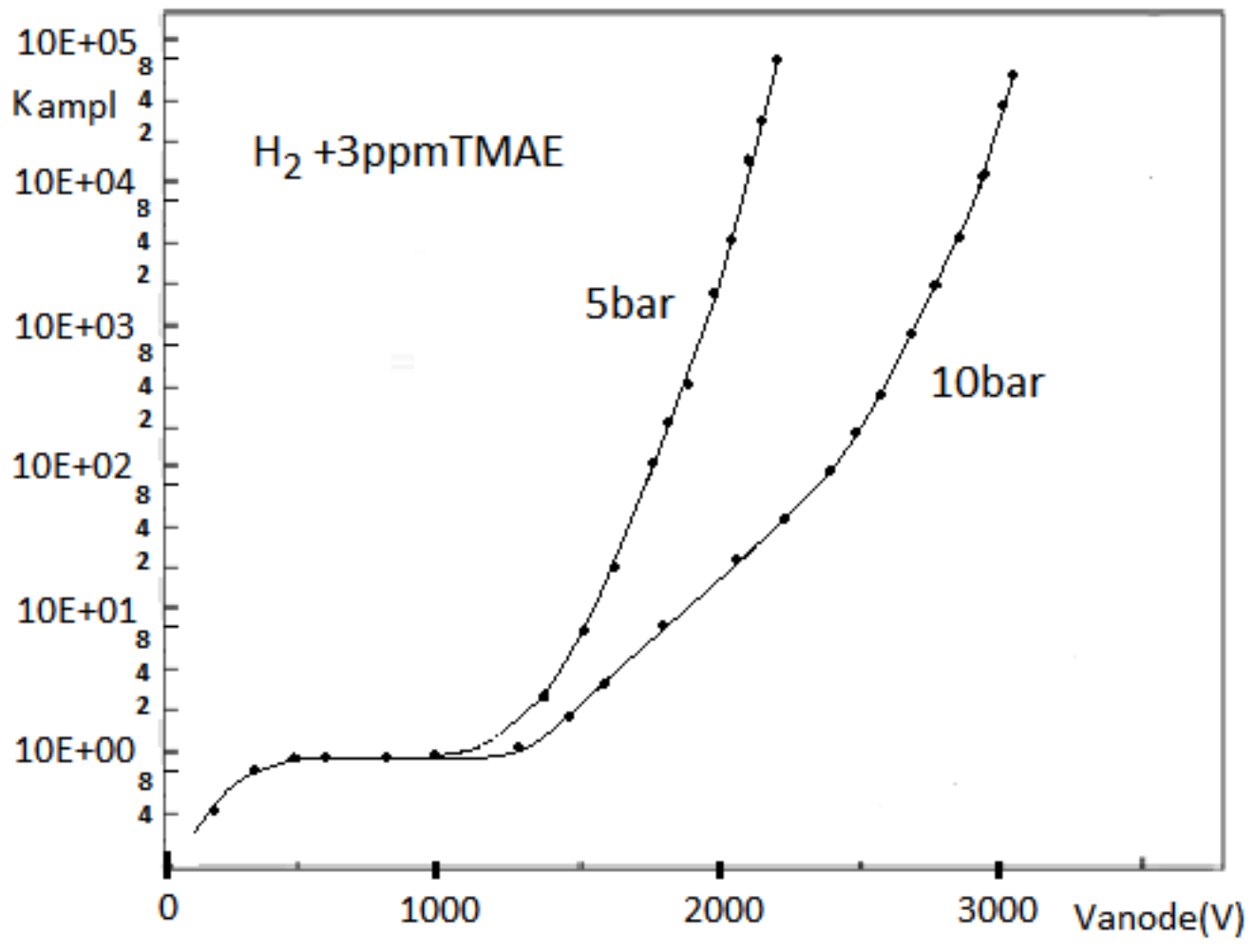

Fig.9. Measured amplification of a system GEM + pin-anode.

a)

| WIMP [GeV/c²] | Nucl. | Enr (keV) |
|---|---|---|
| 0,5 | *H* | 1,91 |
| 1,0 | *H* | 4,30 |
| 1,5 | *H* | 6,20 |
| 2,0 | *H* | 7,65 |
| 2,5 | *H* | 8,78 |
| 3,0 | *H* | 9,68 |
| 0,5 | *Na* | 0,19 |
| 1,0 | *Na* | 0,73 |
| 1,5 | *Na* | 1,57 |
| 4,0 | *Na* | 9,07 |

.

b)

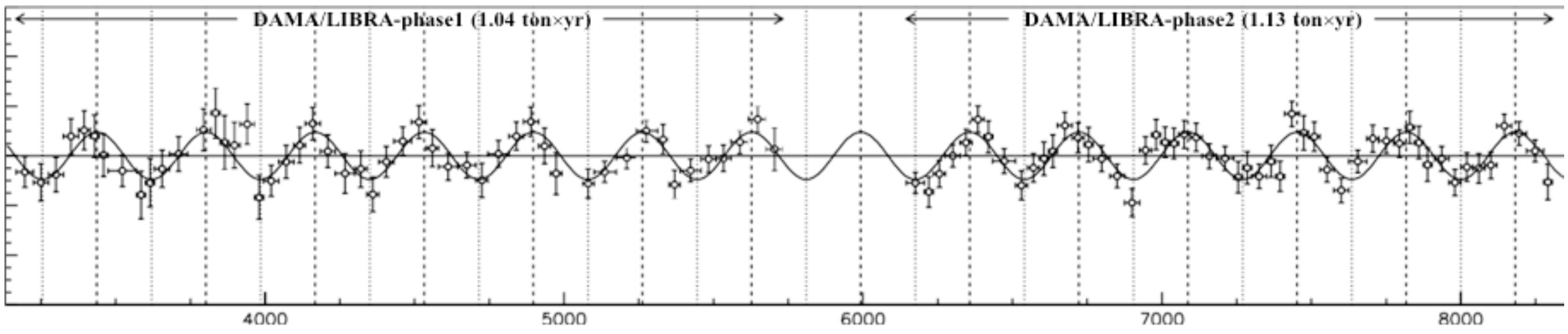

Table1. Maximum calculated nuclear recoil energy $E_{nr}$(keV) as a function of WIMP mass (GeV/$c^2$) for two targets: hydrogen and sodium (a). J.Va'vra has supposed [31] that this effect is explained by low mass WIMP scattered at protons in $H_2O$ molecules(H+), which is contained in NaI crystals at about 1ppm level. Residual rate for single-hit scintillation events in the (2–6) keV energy interval in DAMA/LIBRA–phase1[28] and DAMA/LIBRA–phase 2[29](b).

**6. Chambers with Xe+$CF_4$ (1:1) gas mixture.**

**6.1 Nanosecond timing scintillation chamber with mixture Xe+$CF_4$ filling. The ratio S1/S2 for β and α-particles.**

Recently we have investigated a scintillation signal (S1) and a ratio scintillation to ionization signal (S1/S2) for β and α-particles on prototype of fast chamber (see Fig.10) with mixture Xe+$CF_4$ (1:1) filling at a pressure of 10 bar. This chamber was irradiated with α (239Pu) and β (63Ni) particles. The scintillations signals (S1) were measured separately of photomultiplier (PMT-85) with fast shifter (OB-205). The ionizations signals (S2) were measured on anode of chamber. The addition in Xe of $CF_4$ and suppression of long triplet component of signals (27 ns) ensures the detection of scintillation singlet signals with high speed (1ns).

A shifter OB-205 has a maximum sensitivity range of 185 nm and converts with high efficiency of UV-light in visible light (420 nm). And also he have fast luminescence lifetime (~ 1ns) and high photoluminescence quantum yield (99%). The measurements S1 signal used a fast amplifier and an oscilloscope Le Croy-232. Electronegative impurities $O_2$, $C_2F_4$ and $C_3F_8$ were removed from the gases an a multistage purification system to a level of $10^{-8}$ $O_2$ equivalent (0,01 ppm). The entire system (chamber+ gas system) was checked by the "ISTOK" gas analyzer [32] for the presence of known electronegative impurities, which were not detected as a result.

For α-particles, the ratio A=S1/S2 was 0.63 and B=S1/S2 for β-particles 25.The ratio beta to alpha was B/A=25/0,63=40.

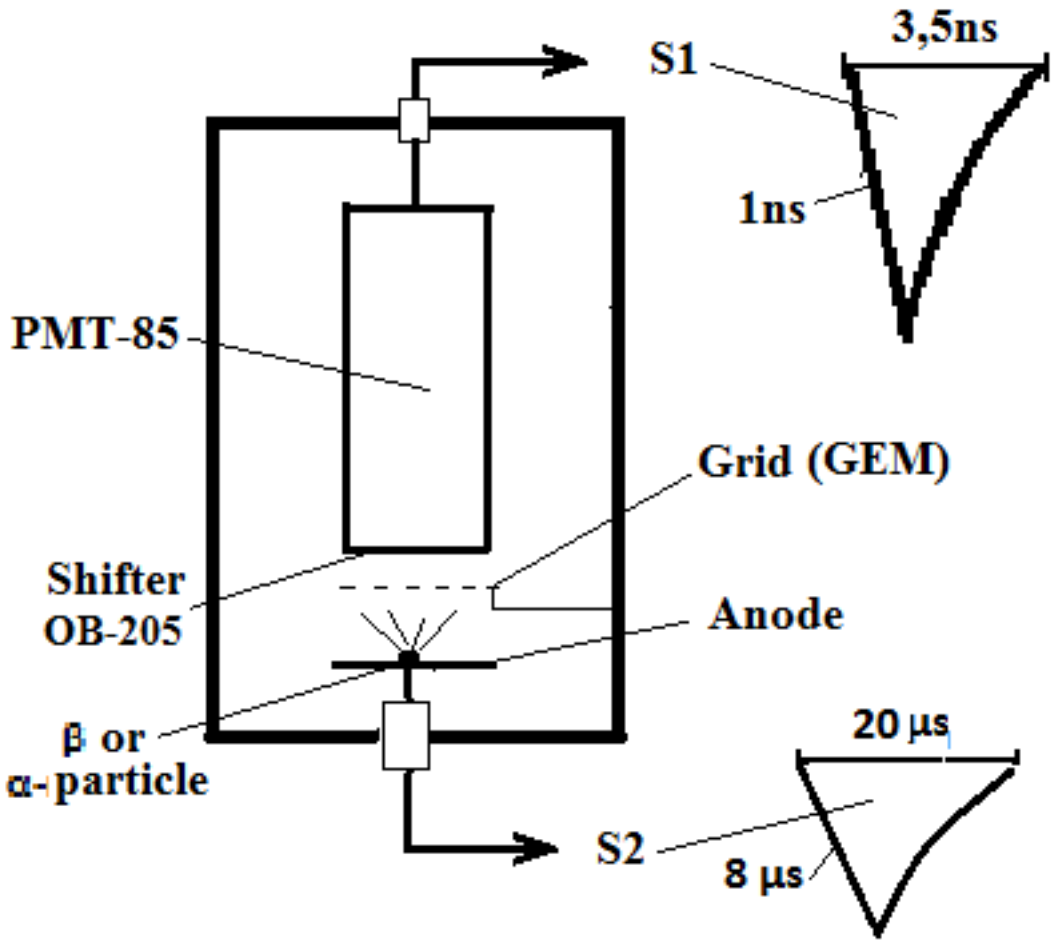

Table 2.

| U(V) | 1000 | 1100 | 1200 | 1300 | 1400 | T(ns) |
|---|---|---|---|---|---|---|
| Xe+CF4 | 200 | 500 | 900 | 1300 | 1800 | 1ns |
| CF4 | 56 | 120 | 200 | 400 | 800 | 2ns |
| Ar+CF4 | 15 | 25 | 70 | 120 | 290 | 3ns |
| Ne+CF4 | 20 | 70 | 180 | 275 | 620 | 6ns |
| Xe | 600 | 1000 | 2200 | 4500 | 7000 | 27ns |

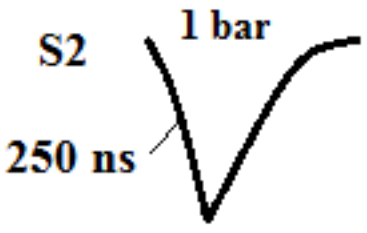

Fig.10. Left: prototype of fast chamber with mixture Xe+CF4 (1:1) filling. Right: amplitude of scintillations signals (mV), which measured of photomultiplier for α-particles on different gases at pressure 10 bar. U (V) - voltage of PMT-85 (1000V – 1400V). The potential applied to anode is Va = +400V, the gap of anode – grid is 8mm.

For mixture Ar+$CF_4$ (1:1) at a pressure 10 bar, the speed is ~3 ns. For mixture Ne+$CF_4$ at a pressure 10 bar, the speed is ~6 ns (see Table 2). The ratio of scintillation singlet signals α-particles to β-particles was U(α)/U(β)=2.To get a electron multiplication factor ($10^3$) and large electroluminescence signals, the chamber instead of a grid is set to WGEM [9]. And to get a large electron multiplication factor ($10^4$- $10^5$) and large ionization anode signals, the chamber instead of a anode is set to system GEM+pin-anodes [16, 17, 21, 25].

**6.2 The sours of electrons with energy 1-100 eV. The study of chamber with hole + pin-anode with Xe + CF4 filling at irradiation with α and β particles**.

The measurements were taken in the pin-anode chamber (Fig. 11 left) [20]. The chamber was used in two operating modes at cathode potentials Vc = 0 and –500 V. In both cases, cathode C2 was grounded. The results obtained thereby are presented in Fig.12. The wide peaks observed when α and β events were detected at Vc = 0 can be attributed to clouds of photoelectrons due to primary scintillations on particle tracks and to photoelectrons produced by photons from avalanches at the pin anode, as well as to ionization electrons from tracks of α and β particles (Fig.12 a). Since the concentration of photosensitive dopant CF4 was high, all three peaks merged into a single wide peak, by contrast to the spectrum from the mixture of Ar + 40 ppm C2H4 in which these three peaks are recorded separately (see section 5.3).

Apparently, CF4 photoionization takes place in the mixture of Xe+ CF4: hν ($Xe_2$*)→ CF4 → $e^+$ + F + $CF_3^-$. When a negative potential of –500 V is applied to chamber cathode C1, all photoelectrons gather on chamber cathode C2, and only ionization electrons from α and β particles are detected at the pin anode. Figure 11 (right) present the multiplication factor of ionization electron at the pin-anode in the chamber from α and β particles tracks as a function of the anode potential in the mixture Xe+CF4 (1:1) at pressures of 1 and 10 bar. In these measurements, voltage Vc = –500 V was applied to cathode C1, and cathode C2 was grounded. For pressure 1 bar the maximum electron multiplication factor equal to 3× $10^4$ was obtained for β particles and Kmax = 3× $10^3$ was obtained for α-particles.

The use of CF4 dopant in noble gas with the aim of increasing the electron drift velocity was described in numerous papers [20]. Our results have demonstrated that CF4 is a photosensitive dopant for Xe. As a result, it is possible to detect scintillations in a chamber filled with a mixture of Xe+CF4 with a 100% geometrical efficiency, with is required in the experiment of the search for DM in the Universe for development of detectors with a high mass and complete suppression of background due to $Kr^{85}$ and external **γ**-rays. As well fluorine-19 ($^{19}F$) has a large spin-dependent WIMP-proton cross-section [35, 36, 37].

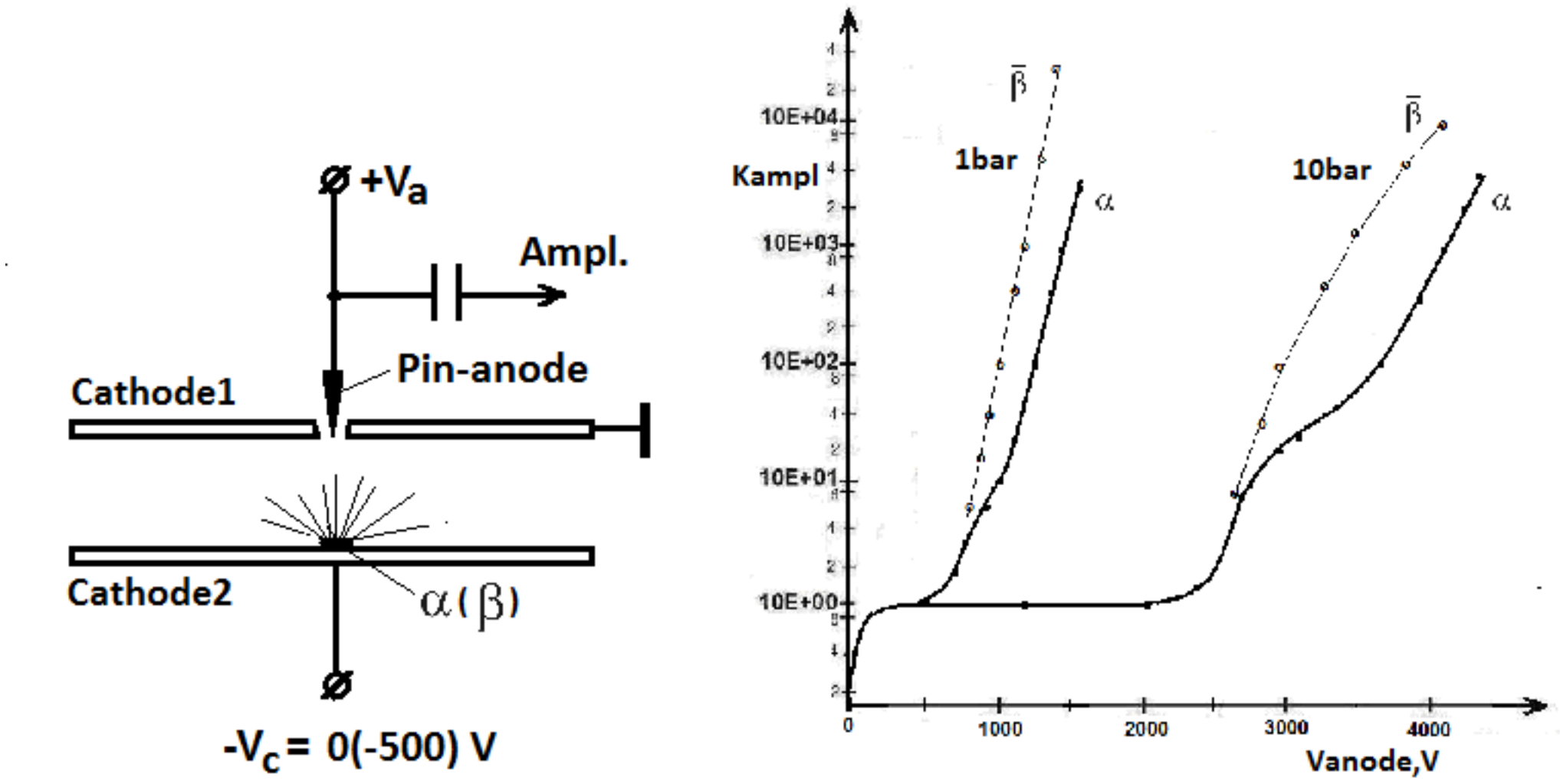

Fig. 11 Left: the chamber with Xe+CF4 (1:1) filling [20]. Right: the multiplication factor (MF) of ionization electron at the pin-anode in the chamber from α (239Pu) and β (63Ni) particles tracks as a function of the anode potential (0 - 4500V).

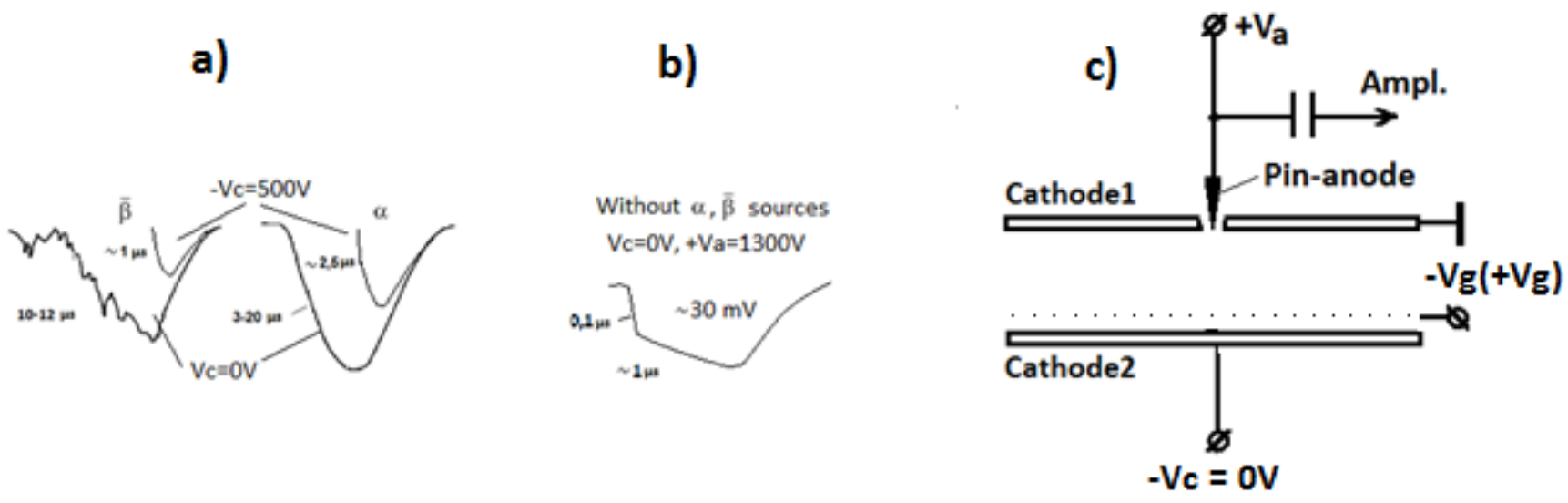

Fig.12.The signals of chamber: (a) at a pressure of 10 bar and irradiation with α and β particles, Va = 3000-4000 V, -Vc = 0(-500 V); (b) The form signals of thermos electrons without α and β sources, Va = 1300V, -Vc= 0V; c) The sours of electrons with energy 1-100 eV, -Vc = 0V, Vg- potential of grid: -10V (electron source closed), +1V to + 50V (electron source open).

In Fig.12b we present the signals of chamber with the mixture of Xe + CF4 (1:1) at a pressure 1 bar without α and β sources. The multiplication factor was equal to 3× 104 (see Fig.11 right). According to the authors, we observe emission electrons from the cathode surface in weak fields of about 10-50 V/cm, which overcame the potential barrier due to their thermal energy. With a gas amplification factor of 3× 104 and average amplitude of signals equal to 5-50 mV the electron energy is about 5-50 eV. The volume of the chamber is thus converted into a low-energy electron source of 1–100 eV. The intensity of electron emission was about 10 per second from 1 $cm^2$ of the cathode surface. At a cathode2 potential of -500V, the electron energy increases and their intensity decreases due to collection at the cathode1. The proportional signals at the anode from single electrons illustrate the Penning effect (see Fig.12b). Figure 12c illustrates construction the sours of electrons with energy 1-100 eV.

**7.** Search **for spin-dependent WIMP-nucleon interactions.**

**7.1 Double-phase xenon chamber with a system GEM +pin-anodes.**

In particle physics, the lightest supersymmetric particle (LSP) is the generic name given to the lightest of the additional hypothetical particles found in supersymmetric models. In models with R-parity conservation, the LSP is stable; in other words, it cannot decay into any Standard Model particle, since all SM particles have the opposite R-parity. There is extensive observational evidence for an additional component of the matter density in the universe, which goes under the name dark matter. The LSP of supersymmetric models is a dark matter candidate and is a weakly interacting massive particle (WIMP).

From the elementary particle physics in the framework of the standard Bing Bang nucleosynthesis model one infers that DM consists manly of WIMPs: massive neutrinos, axions and particles predicted by SUSY. The most probable candidates for the WIMP are the neutralino, predicted by supersymmetric theories (SUSY) [35].

The direct method for WIMP search consists in detection for their elastic scattering on detector nuclei. The WIMP interaction probability in detector can be represented in a form: $R = a_p W_p^2 + a_n W_n^2 + a_0 V^2$, where the first two terms determine the spin-dependent (sd) scattering and the last one is the spin-independent (si) scattering [35]. The ratio of the number of spin-dependent WIMP scattering Rsd to the number of spin-independent scattering Rsi can be represent in a form:

$$Rsd / Rsi = \eta_A \cdot \eta_{SUSY},$$

where $\eta_A$ is determined by nuclear structure, and $\eta_{SUSY}$ – by neutralino-quark interaction in SUSY model.

The dependence of $\eta_A$ from atomic number A for nuclei ($^1H$, $^3He$, $^{19}F$, $^{73}Ge$, $^{127}I$, $^{205}Tl$ and others) with nonzero spins is shown in Fig. 13a (left) [35]. The dependence Rsd (A)/ Rsi ($Ge^{73}$) on neutralino mass are shown in Fig. 13a (right) for nuclei $^{19}F$ and NaI [36]. One can see that more strong restrictions on spin-dependent part of WIMP interactions can be obtained in experiment with $^{19}F$ ($CF_4$, LAr+$CF_4$, LXe+$CF_4$) as compared with other nuclei [37]. The measurements are especially attractive in region of WIMP mass 8 GeV ≤ $m_x$ ≤ 14 GeV, where Rsd > Rsi.

At present time a great experiments [24, 26-28] are carried out for WIMP search with detector containing nuclei LAr, LXe, NaJ, Ge, the spin-independent (coherent) scattering for which is large [29]. One can see, that sensitivity of these experiments for spin-dependent scattering is 10-100 times less of expected effect as distinguished from coherent scattering experiments (see. Fig.13).The proposed in this work experiments increases the sensitivity of spin-dependent measurements to the point of the expected effect [23].

In this context, in our experiments [21, 25, 39] as another filling of the chambers for search of low-mas WIMP (<10 GeV/$c^2$) and solar axions with spin-dependent interaction with deuterium ($D_2$), $^3He$, $^{19}F$ and $^{21}Ne$ we propose to use the mixtures: $^{21}Ne$ + 10%$H_2$, $D_2$ + 3ppmTMAE, $^3He$ + 3%$CH_4$ at pressure 10-17 bar. And in our experiments [16-18] with liquid gases mixtures is used the mixtures with $^{19}F$: LAr + $CF_4$ and LXe + $CF_4$ (see Fig. 14). The relative scintillation light outputs for investigated gases evaluated by an distribution area is shown in Table 3.

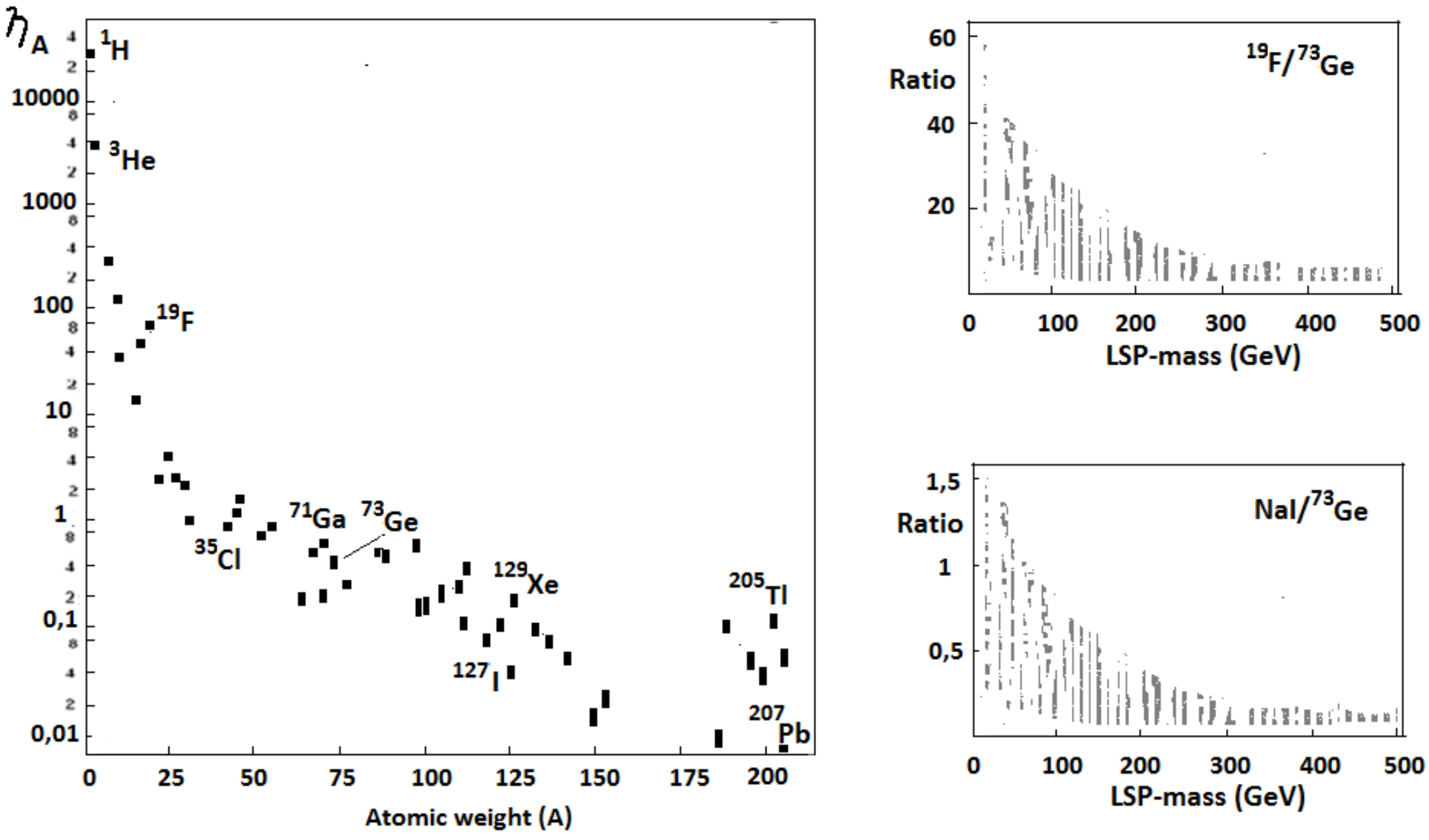

Fig. 13a. Left: The dependence of nuclear factor $\eta_A$ from atomic number A for the nuclei with nonzero spins [35]. The high of symbols presents the change of $\eta_A$ in a WIMP mass interval 10 GeV ≤ $m_x$ ≤ 500 GeV. Right: The dependence of Rsd (A)/ Rsd ($^{73}$Ge) from WIMP mass for $^{19}$F and NaI [36].

| Gas | Xe gas | LXe | Xe+$CF_4$ | LXe+$CF_4$ | L$CF_4$/$CF_4$ | LAr+$CF_4$/Ar+$CF_4$ |
|---|---|---|---|---|---|---|
| Scintil. output | 1 | 1 | 0,3 | 0,16 | 0,5 | 0,5 |

Table 3. The ratio of relative scintillation light outputs for different gases mixtures [44].

In this section, we describe a double-phase xenon chamber with a system GEM +pin-anodes and system of photomultipliers. Fig. 13b shows a design of the chamber.

To suppress the β, γ and n0 background, we propose a comparison of scintillation singlet signal (S1) and ionization signal (S2) for every event is suggested. The addition in Xe of $CF_4$ and suppression of long triplet component of signals (27 ns) ensures the detection of scintillation signals with high efficiency and provides a complete suppression of the electron background. The singlet component S1 of the signal (1ns) is determined by the quenching factor (QF) of scintillation signals [47].

The scintillation intensity at the same particle energy is determined by the ratio 10: 5: 1 for electrons, protons and alpha particles [48]. That is, on the tracks of alpha particles, only ~10% of their energy is spent on the singlet scintillation component. On the contrary, only ~10% of the energy of beta particles is spent on ionization of the component, and ~90% is the singlet scintillation component of the signals.

For alpha particles, the ratio A=S1/S2 was 0.63 and B=S1/S2 for beta particles 25. The ratio beta to alpha was B/A=25/0,63=40 for 50%Xe + 50%$CF_4$ gases mixture ( see data of section 6.1 and 6.2).

It is possible that in mixture Xe + CF4 (see section 6.1) the singlet component of signal (1ns) for beta particles with spin (J=1/2) is due to spin-dependent interaction between beta particles and molecules gases with non-zero spin ($^{19}F$, $^{129}Xe$, $^{131}Xe$). This assumption is indicated by a comparison of S1 and S2 for mixture 50%Xe + 50%$CF_4$ and mixture Ar + 40ppm C2H4 with zero-spin (see section 5.3). In the mixture Ar + C2H4, the average ratio of S1 to S2 does not exceed ten. The long triplet ($I_t$) components of the signals due to the formation of dimers are suppressed. In general, they do not depend on the spins of the particles, because they are determined by long secondary processes of gases dimers. For alpha particles with zero-spin the singlet component of signal (1,5ns) is not due to spin-dependent interaction.

An important fact for understanding the processes is that the ratio between the singlet ($I_s$) and the triplet ($I_t$) intensities in Ar is equal to $I_s/I_t$=0.3($e^-$), 1.3($\alpha$) and 3(nr) [51]. Another explanation for the measurement results is that the dimers are completely suppressed in Xe + CF4 and only partially in Ar + C2H4.

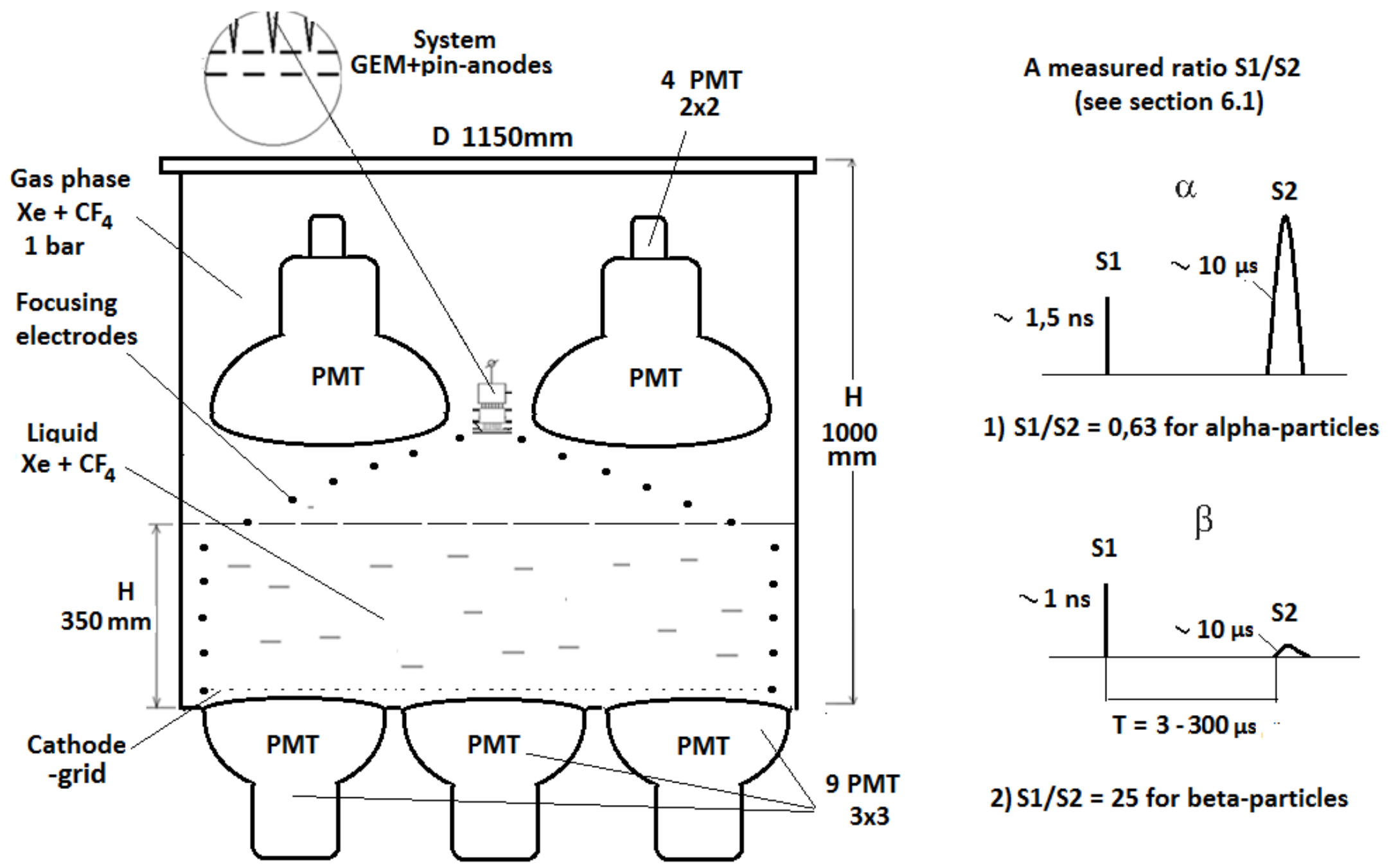

Fig. 13b. Double-phase Xe + $CF_4$ chamber with a system GEM +pin-anodes and photomultipliers.

The delay time between S1 and S2 is T = 3-300 µs, because: the drift time of electrons is T = 0-300 µs in LXe + $CF_4$, and the drift time electrons is T = 1-3 µs in gases phase 50%Xe + 50%$CF_4$ [44].

### 7.2 The double-phase chamber with $LCH_4$ + TMAE mixture for search of spin-dependent WIMP-nucleon interactions.

In work [46] it is shown that the absolute lower bound for the rate of direct DM detection is due to the spin-dependent WIMP-nucleon interaction, and a new-generation experiment aimed at detecting DM with sensitivity higher than $10^{-5}$ event/(kg · day) should have a nonzero-spin target to avoid missing of the DM signal. In work is claimed that for targets with spin- nonzero nuclei it might be the spin-dependent interaction that determines the lower bound for the direct detection rate when the cross section of the scalar interaction, which is usually assumed to be the dominant part, drops below $10^{-12-13}$ pb particular, from this work one can see that all fluorine-containing targets (LiF, $CF_4$, $C_2F_6$, and $CaF_2$, etc.) have almost the same sensitivity to both the SD and SI WIMP-nucleus interactions.

Among all materials considered a detector with a $^{73}Ge$, $^{129}Xe$, or NaI target has better prospects to confirm or to reject the DAMA result [29] due to the largest values of the lower bounds for the total rate R(10, 50) > 0.06 -0.08 events/(kg day). If, for example, one ignores the SI WIMP interaction, then all materials have almost the same prospects to detect DM particles with the only exception of $CH_4$ (see Fig. 14). The results obtained are based on previous evaluations of the neutralino-proton (neutron) spin and scalar cross sections for the neutralino masses $m_\chi < 200$ GeV/$c^2$.

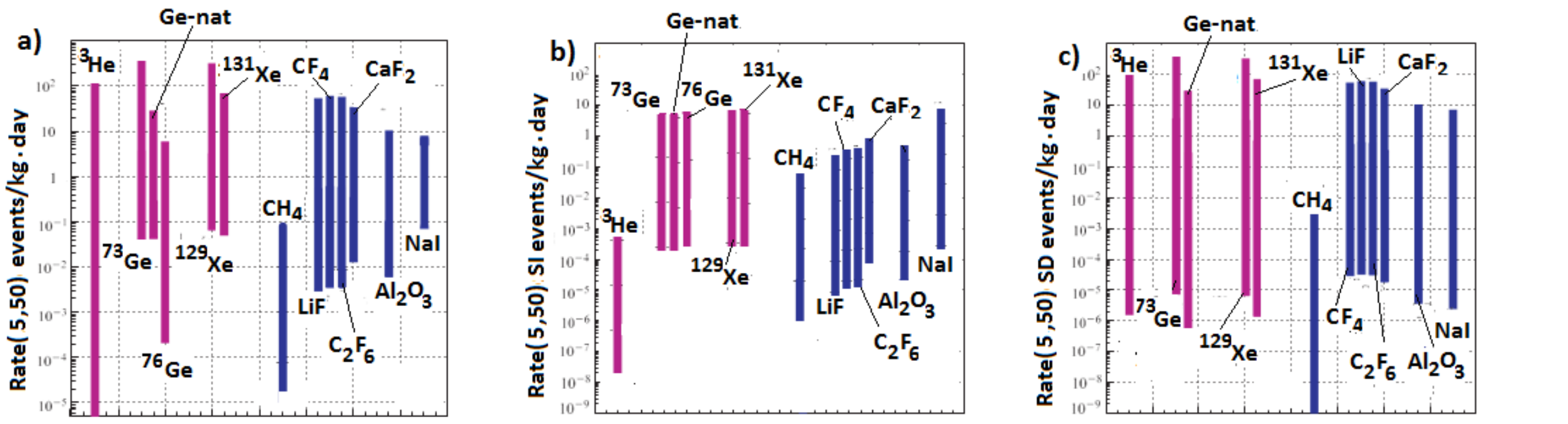

Fig. 14: a) variations of expected event rates, *R* (5, 50), for a number of targets followed from the DAMA-allowed cross sections $\sigma_{SD}$ and $\sigma_{SI}$. Targets with nonzero-spin nuclei from the odd-neutron (odd-proton) group model are given in the left (right) part of the figure; b) variations of expected spin-independent contributions to the event rate, R (5, 50)SI, in a number of targets followed from the DAMA-allowed cross sections σSD and σSI; c) the same as in b, but for the spin-dependent contributions *R* (5, 50)SD [46].

In this context, in our experiment [15] with liquid $CH_4$ as another filling of the chamber for search of low-mas WIMP (<10 GeV/$c^2$) we propose to use the mixture $LCH_4$ + 40ppmTMAE (see Fig. 15).

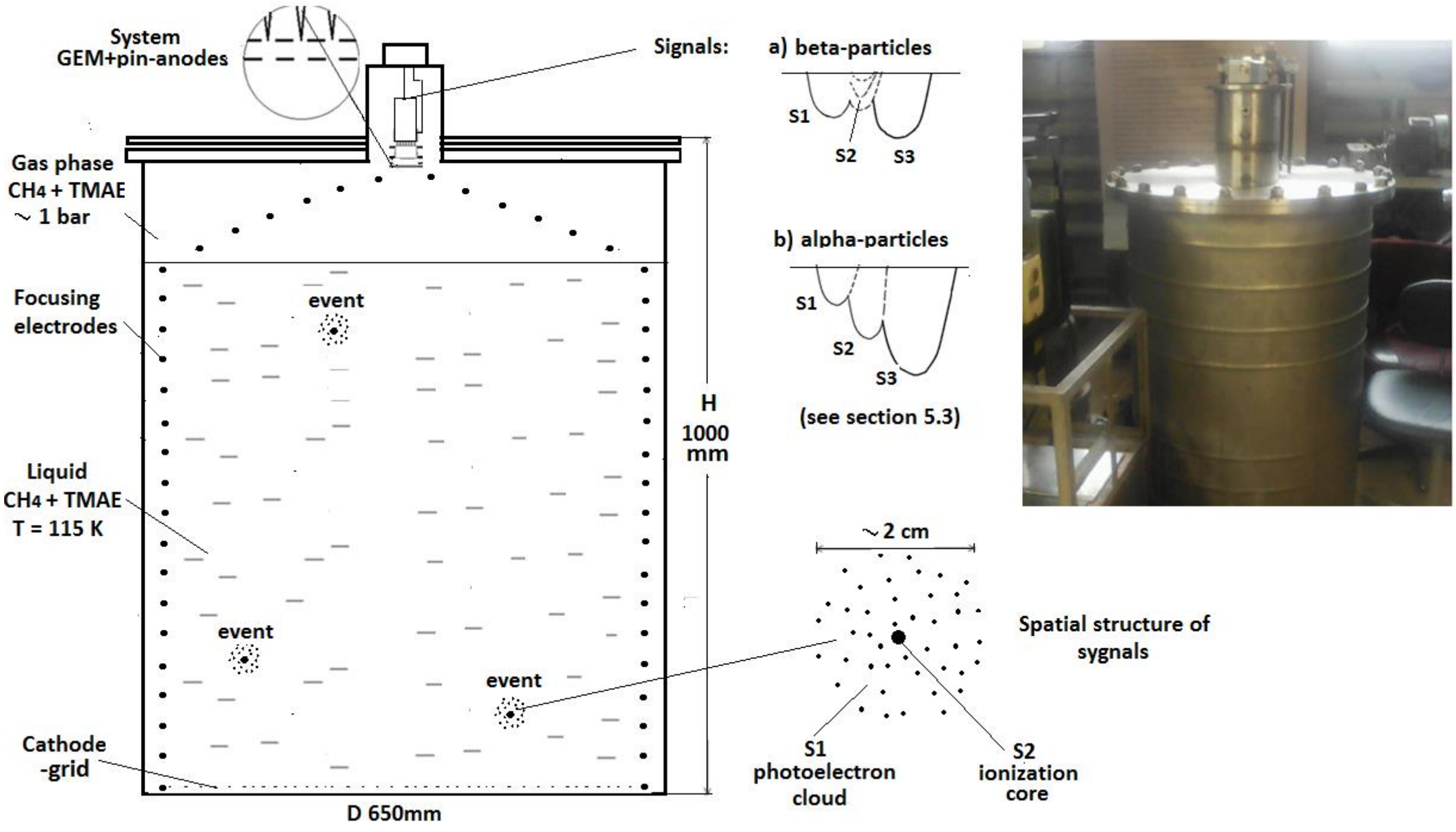

Fig. 15. A liquid-methane ionization chamber with a system GEM + pin-anodes for search of spin-dependent WIMP-nucleon interactions.

The body of the chamber is made of titanium. The cathode-grid of the chamber is immersed in liquid methane. The layer of liquid methane above the cathode is equal to 800 mm. The temperature of liquid methane is equal to 115 K and the pressure of gaseous methane over the liquid methane is equal to 1,3 bar for this temperature. The system GEM + pin-anodes is placed in gaseous methane. The addition in liquid $CH_4$ of photosensitive TMAE [19] and suppression of triplet component of scintillation signals ensures the detection of scintillation signals with high efficiency and provides a complete suppression of the electron background (see section 5.3).

## 8. Time-projection chamber with high dE/dz and energy resolutions.

It is known that multi-wire proportional chambers (MWPC) used in time-projection chambers (TPC) fail to provide good energy resolution [34]. This is mainly due to the fact that multiplication of ionization electrons occurs in chamber regions with different electric field intensities resulting in variation of the multiplication factor. Additionally, the MWPC features a low dEI/dz resolution in the drift direction of the ionization electrons due to slow motion of positive ion clouds from anode wires.

In our paper [38], we describe a TPC with both high energy and dE/dz resolutions. Fig. 16a shows a design of the chamber. The chamber contains the cathode 1 with a 4-mm-diameter collimating opening, behind which a 239Pu source with a $10^5$ flow rate is installed at a distance of 10 mm. The rings 3 provide a uniform drift field. The MWPC 5 used for electron multiplication contains the cathode MWPC wound with a 2-mm pitch of a beryllium bronze 100-µm-diameter wire. The anode is winded of a 20 µm diameter W + Au wire with a 2 mm pitch. The MWPC gap is 2 mm. In projection, the anode wires are exactly in the middle between the cathode wires. The chamber contains cathode 6 and anode 7 of the measurement gap. The maximum electron multiplication factor for α-particles is $\sim 10^3$. To get the total electron multiplication factor ($10^5$), the chamber is set to MGEM 4 with the etching of holes [14].

Due to the photon mechanism, when the avalanche evolves in the anode MWPC wire, an electron charge proportional to the initial ionization charge appears in the transfer gap between the electrodes 5 and 6. This charge drifts towards the measurement gap between the electrodes 6 and 7. Electrodes 6 and 7 are winded of a 100 µm diameter wire with a 1 mm pitch. The width of the gap 5-6 is 4 mm. The electric field intensity in this gap is insufficient for evolving avalanches and the gap records the electrons without multiplication, i.e., in the induction operation mode of the ionization chamber.

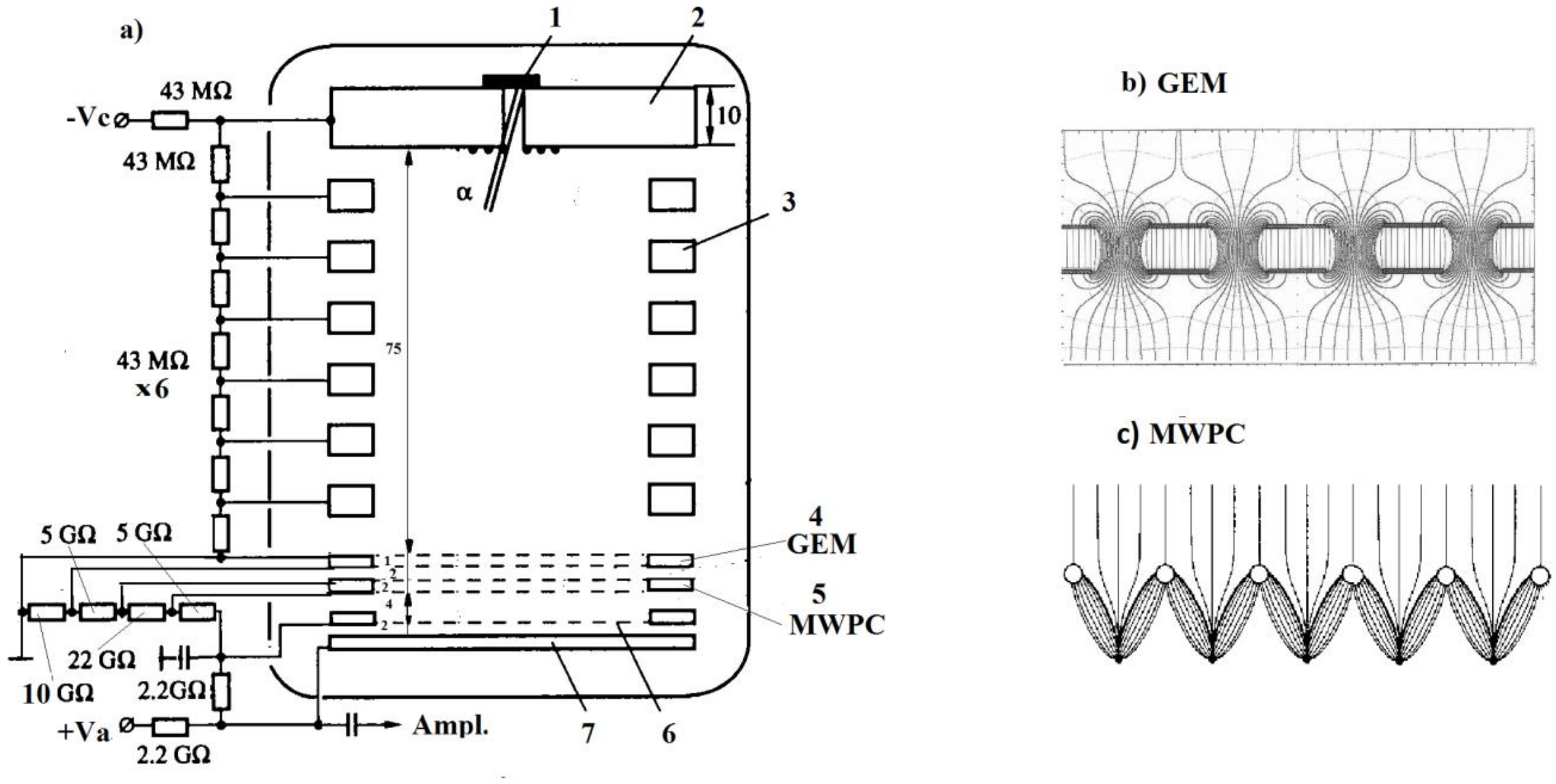

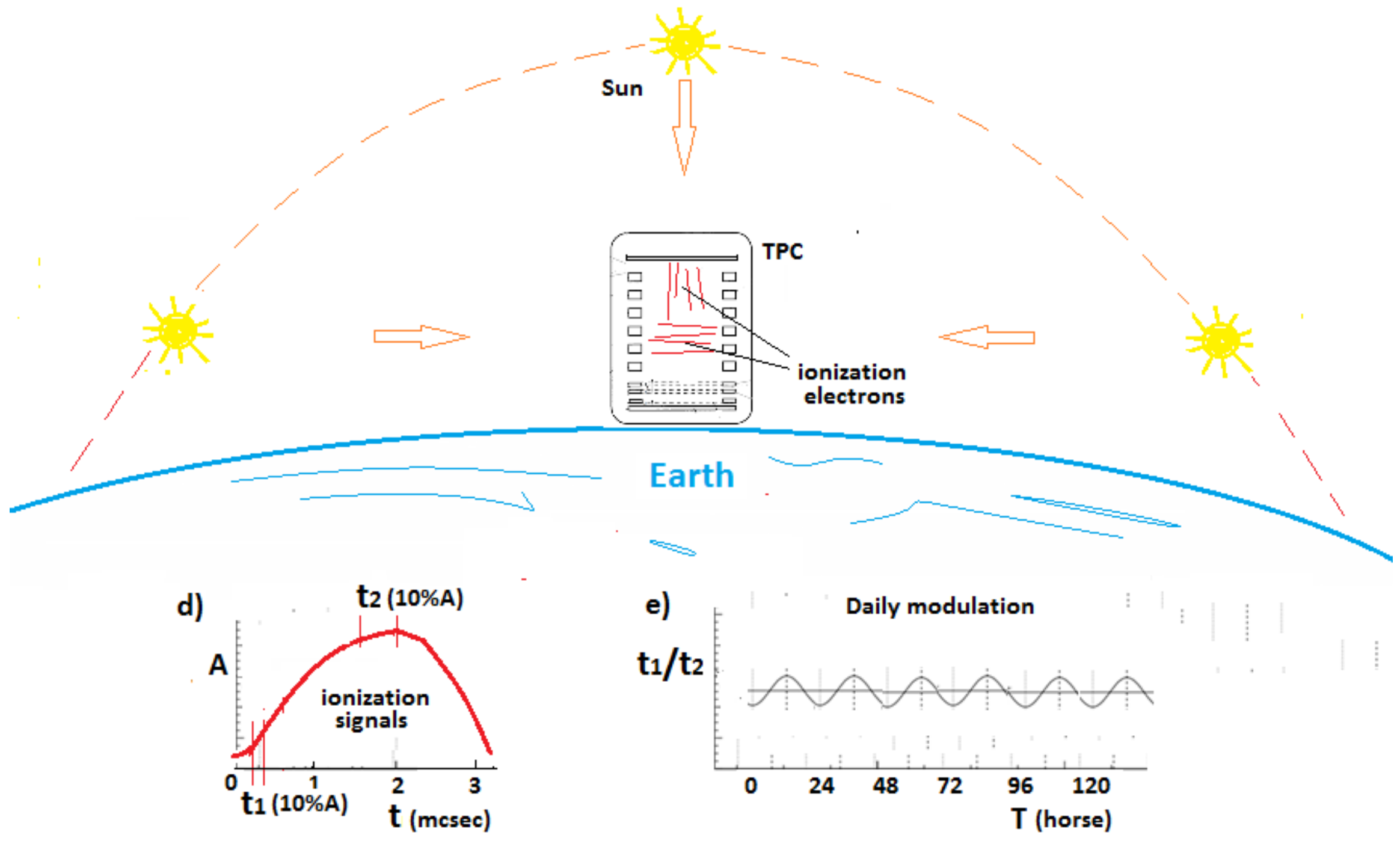

Fig.16. Design of time projection chamber (TPC) with MWPC and GEM (a), electric field structure in the GEM (b), MWPC (c), signals of TPC (d) and daily modulation of signals (e).

Fig. 16 (b, c) shows the electric field structure in the MWPC and GEM gaps. Since the electric field intensity in the MWPC gap is much higher than in the drift gap and due to the symmetric location of the cathode relative to the anode wires, the electric field lines, when passing from the drift gap to the MWPC gap, concentrate into a narrow beam, which is orthogonal on the average to the anode plane and sufficiently uniform. As a result, the electron multiplication occurs in a more homogeneous electric field than in usual MWPCs, in which the electric field lines meet the anode wires at different angles in different MWPC regions and have different average densities.

In the process of its evolution, the avalanche discharge involves about 180° of the anode wire in the azimuth. In our case, conditions for the evolving discharge are the same for all ionization electrons within the drift gap, because the field distributions on all wires are identical, and the fields are more homogeneous (see Fig. 16 d, e).

Since, for usually utilized mixtures, the electron drift speed in the measurement and drift gaps is $\sim n \cdot 10^6$ and $\sim n \cdot 10^5$ cm/s, respectively, and the measurement gap is small, the calculated dE/dz resolution referred to the TPC drift gap appears to be sufficiently high: the chamber resolves the track regions or separate tracks spaced by ~0.2 mm.

In order to determine the energy resolution and signal structure the chamber was filled with an Ar + 0.5% C2H2 + 10%CH4 mixture at a total pressure of 3,5 bar. When recording a-particles from a 239Pu source, the energy resolution was 8.1% (half-intensity half-width). The front of signals was 2 µsec. (Fig.16d).

Three anode grids with 0,2mm pitch and 0,2mm gaps may be placed in the ionization chamber for detection of x, y and 45° x-coordinates. The electrons are moving after multiplication through three anode grids inducting on them the signals. The errors for x, y-coordinates will be equal to 0,2mm and error for z-coordinate will be 0,2mm (not including diffusion errors).

The method allows you to measure the direction of movement of particles. In our work [45] we have study of TPC with the Penning mixture He+3%$CH_4$ filling at a pressure 17 bar for direct detection of solar neutrinos. As another filling is use the mixture $^3He$ + 3%$CH_4$ for search of SD-interactions. As well time projection chamber with the mixture $D_2$ + 3ppmTMAE filling at a pressure 10-20 bar allow to search of spin-dependent interactions of solar axions and deuterium [21, 23, 25]. The chamber can be placed in low background laboratory in low background shielding for search the yearly or daily modulation effects.

Finally, we compare this chamber with existing multistage avalanche chambers, which also provide high energy and dE/dz resolutions and with the MICROMEGAS chambers [7]. The MWPC used in our chamber in place of the avalanche chamber allows it to operate under an increased pressure of the filling gas (up to 10-20 bar), whereas other chambers are intended for operation mostly at pressures lower than the atmospheric pressure. The high resolution (dE/dx, dE/y, dE/z) of the TPC allows you to measure the direction of the flow of solar neutrinos and axions. Directionality of events detecting allows to use only electrons recoiling away from the sun, effectively eliminating most background events.

## 9. Detecting chamber with system GEM+pin-anodes for search of narrow pp-resonances (quarks) at the accelerators in the energy region 150 – 300 MeV.

The existence of narrow peculiarities in a two-proton system was observed for the first time by JINR scientists in nucleon reactions with π-mesons production [40]. By now the statistics has substantially been enlarged by this group and narrow pp-peaks at the level 3-5 standard errors have proved to be present in various reactions within a wide energy range [41]. Analyzing the effective mass distributions in the system of particles np, $\pi^- \pi^-$, $\pi^+ \pi^-$ from inelastic reactions with π-mesons production this group also found the narrow peculiarities, excitation energies of which were just the same as those for pp-system. By now this group has observed narrow $\pi^-$ p-resonances which excitation energies coincide well with those of pp-systems. In work [42] Y.A.Troyan group has observed the existence of narrow pp-resonances in elastic pp-scattering differential cross-section within 116-199 MeV.

The results are compared with the data of various experiments of elastic scattering at the energy region 0,2 - 10 GeV/$c^2$ (Fig. 17).

The experiment with MPWC which detectors the recoil protons with mixture Xe + $iC_4H_{10}$ (1:1) under a pressure 15 torr we proposed in year 1993 to search for narrow pp-resonances (quarks) in the energy region 150 – 300 MeV at the MMF accelerator [43]. In the experiment the external proton beam of the accelerator is used with the energy changed from 160 MeV to 300 MeV, intensity 2 × $10^{11}$ protons per second and energy spread 0,5 MeV (FWHM). The formed proton beam goes through the tube filled with hydrogen (H2) under a pressure of 5 bar (see Fig.13). Recoil protons from the elastic pp-scattering are detected at angles 70° laboratory system for coincidence with gas proportional and scintillation detectors. Detectors adsorb the recoil protons with the energies up to 35 MeV. This experiment requires high speed of ionization electrons collection time and respectively, the detector resolution time is about $10^{-8}$ sec.

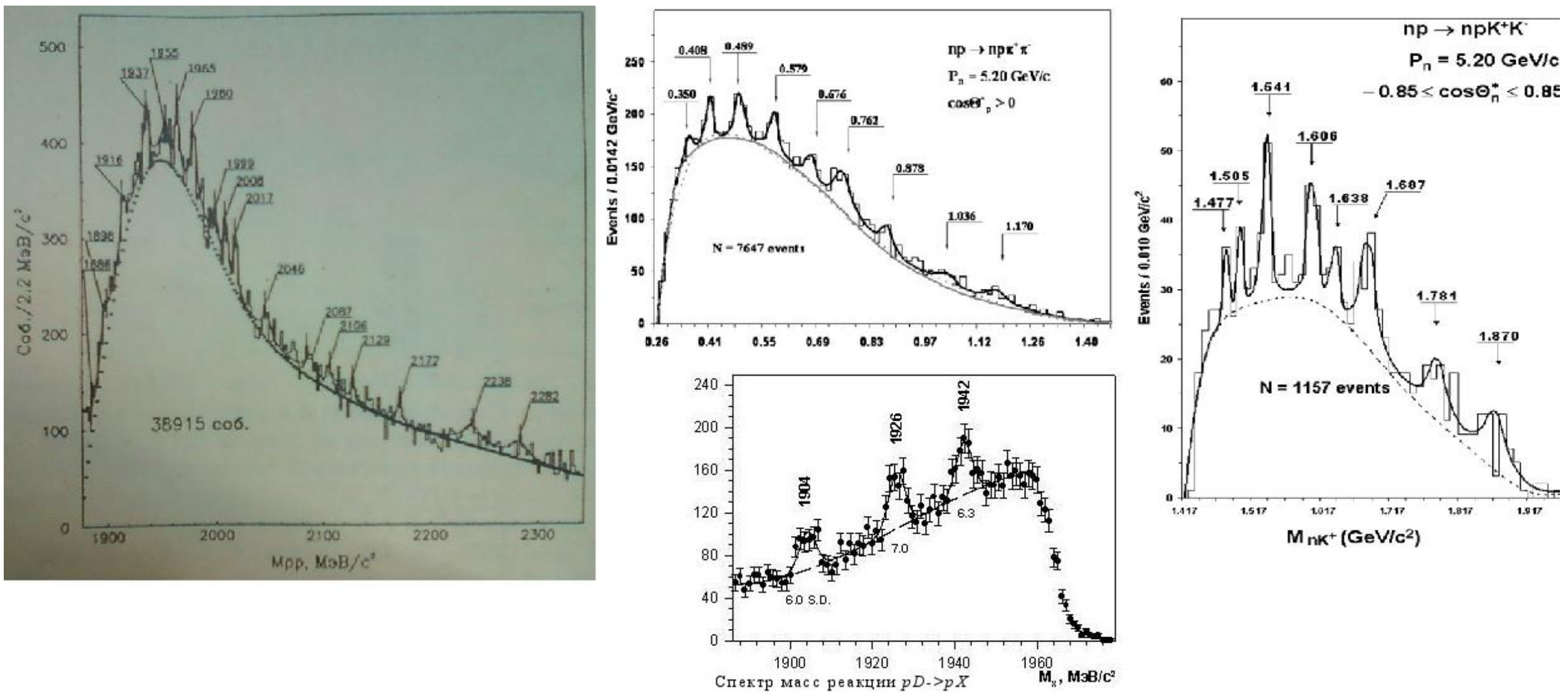

Fig. 17.The results of experiments JINR (Dubna) and MMF (INR, Moscow).

We propose detect the recoil protons with the energies up to 35 MeV on scintillation chamber (see section 6.1) with Xe+CF4 (1:1) filling under a pressure 50 torr and ionization chamber with system GEM+pin-anodes under a pressure 15 torr(see Fig.18). Signals are picked up from the scintillation and ionization chamber by fast amplifier and then are transferred to coincidence and anticoincidence circuits with a resolving time $10^{-8}$ sec. When the signal from the scintillation chamber coincides with that from the ionization chamber the signal amplitude of the latter is registered by the amplitude analyzer. Its readings determine the total number of the events at a certain energy and detection angle ($N_{events}$ + $N_{backgr}$). The ratio $N_{events}/N_{backgr}$ is determined an elastic peak against the background.

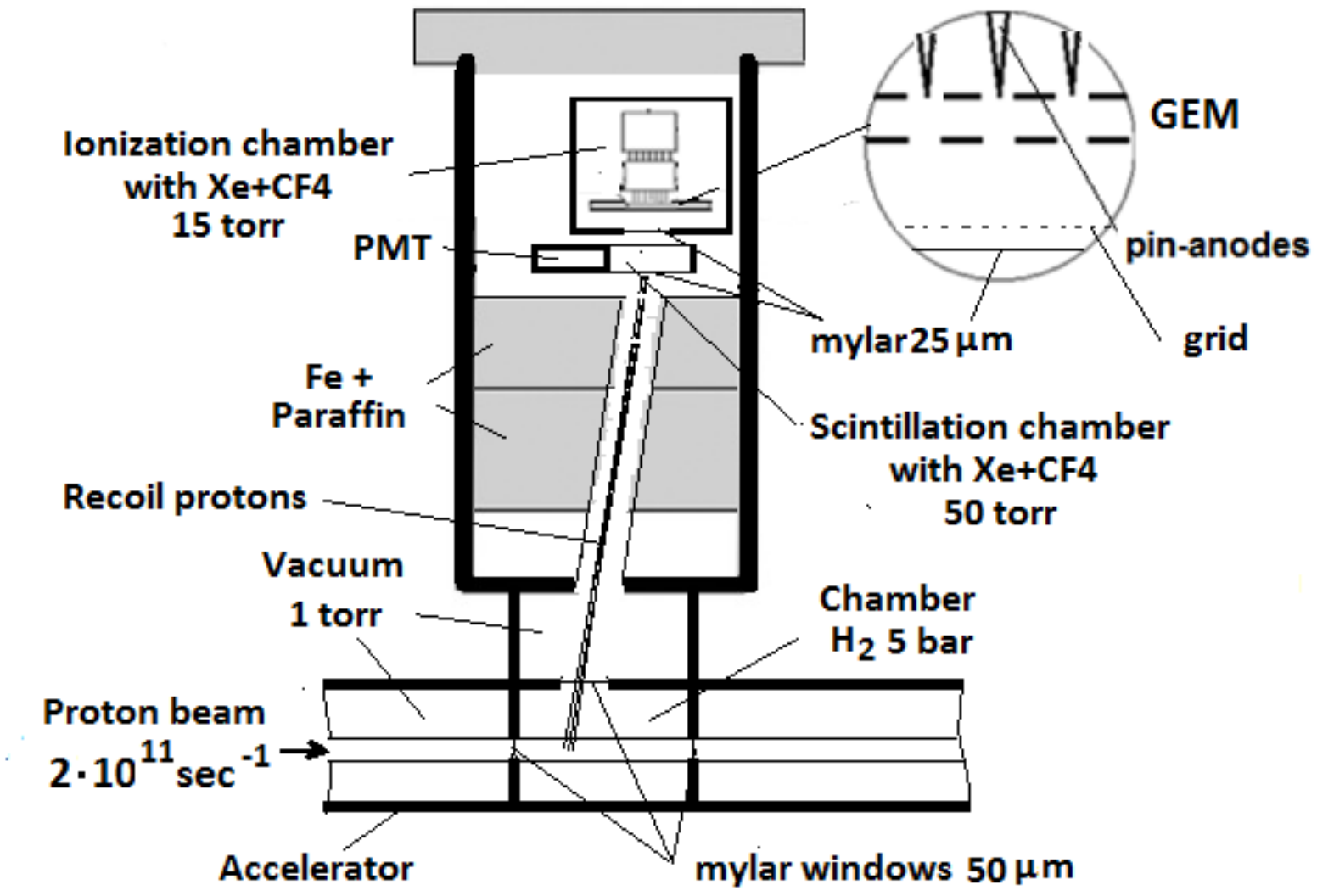

Fig.18. Experimental setup with scintillation and ionization chambers.

We would like to mention that chambers with system MGEM+pin-anodes, as far as the MWPC [1] can have many applications for detecting events in high energy physics experiments.

### 10. Operation principle of MWPC, GEM and system GEM+pin-anode

The operation principle of Multi-Wire Proportional Chambers (a), GEM (b) and system GEM + pin-anode (c) is illustrated in Fig.19. The factors which have allowed us to obtain high electron multiplication factors in the GEM + pin-anode system are as follows:

(1) High electric field strength in the system GEM + pin-anode makes it possible to obtain a big length of electron avalanche and high value of the electron multiplication factor ($10^6$-$10^7$);

(2) Positive ions from the avalanche at the pin are transferred by the electric field, mainly, to the walls of the hole in which the pin is located and, in smaller quantities, towards the ionization electrons being collected at the pin, which rules out the possibility of streamers being developed at the interface

(3) For GEM (see technology b) extraction efficiency decrease at low transfer fields values due to a worst electron extraction capability from the lower side of the GEM [32];

(4) Absence of a plastic insulation excludes the emergence of leakage current and spark breakdown between electrodes. Accidental spark events in such system don't lead to their failure as positive ions quickly move away from breakdown by a strong electric gap field.

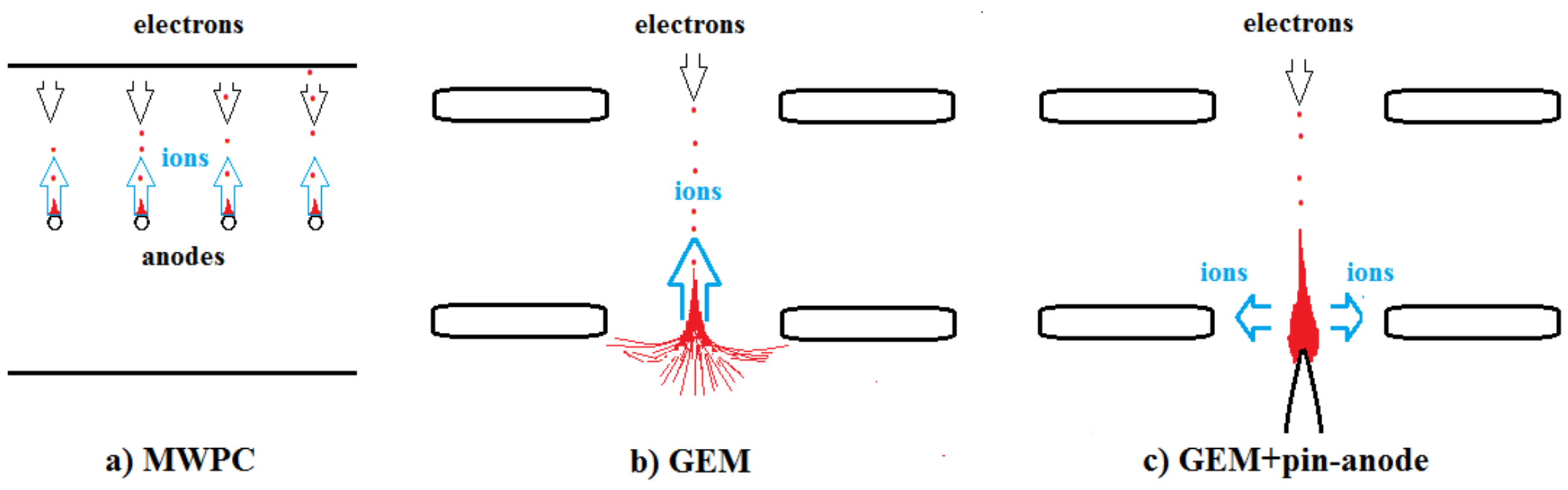

Fig.19. Operation principle of Multi-Wire Proportional Chambers (a), GEM (b) and GEM+pin-anode (c). Electron avalanches are shown for three technologies (a, b, c); red paths are electron trajectories, also the drift of ions is indicated (blue paths).

The leading edges of the signals picked off the pin-anode are 1-3 µs on different gases and their mixtures; however, signal differentiation helps extract the initial parts of the edges 0,2-0,3 µs long. The steepness of the pulse leading edges at the pin-anode is determined by the geometry of the pin-hole geometry and the potential difference between them. The high electric field strength near the pin-anode ensures a fast removal of an envelope of positive ions from pin, which, in combination with the small gaps between the pin and the walls of the hole-cathode (~0,4mm), ensures a rapid decrease in amplitude of the signal induced at the pin-anode by the positive ion envelope; i.e., it guarantees a high steepness of the pulse leading edge (~1µs). The high gas amplification factor in the systems GEM+pin-anode or hole+pin-anode makes it possible to register thermal electrons from the cathode of the ionization chamber with an energy of 1–100 eV. (see section 6.2)

We have also plans for constructing of large scale (150mmx150mm) MGEM + pin-anodes detectors (see Fig.20).

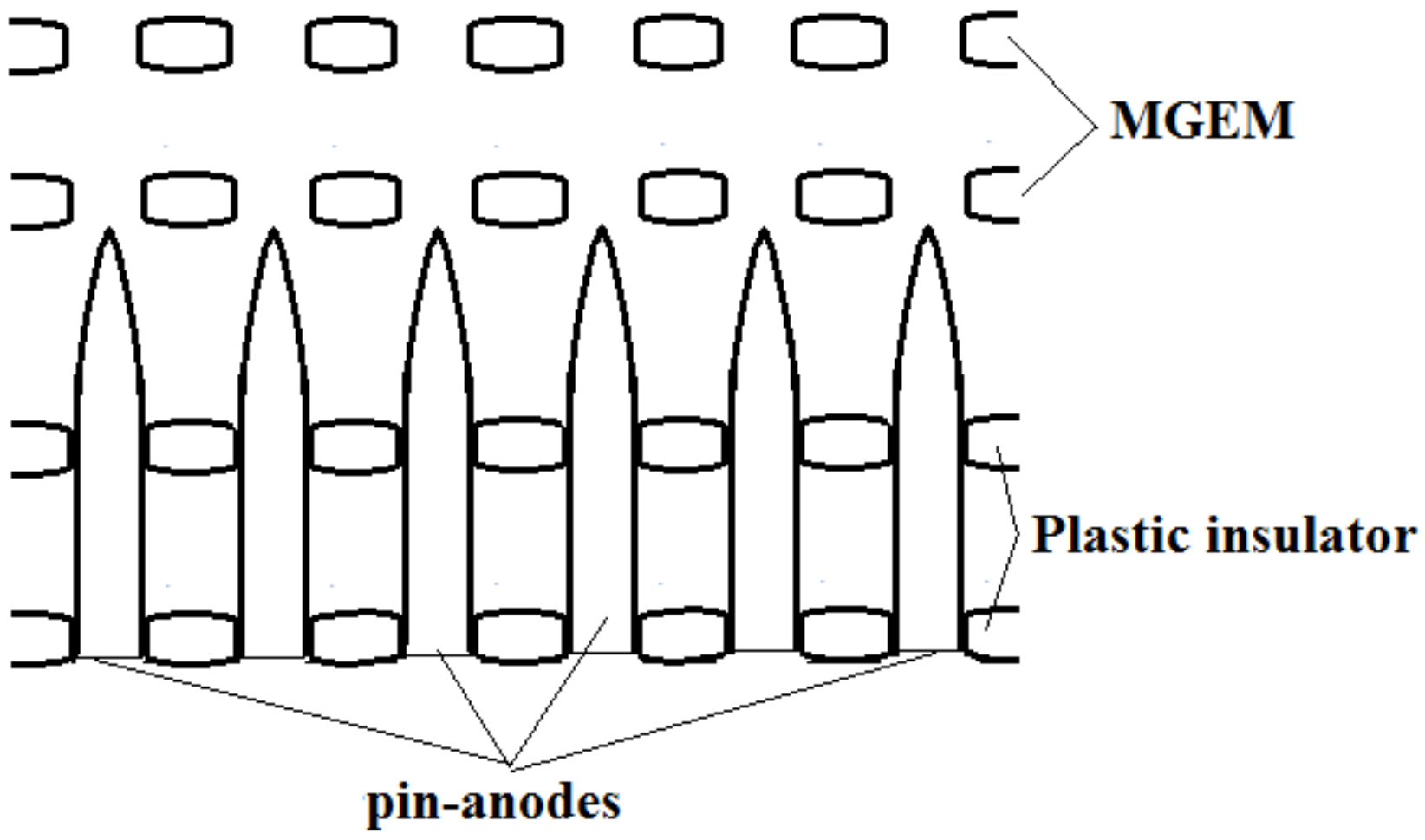

Fig.20. Schematic drawing of system GEM + pin-anodes in which plastic insulator contains of holes for pin-anodes. The metal electrodes (MGEM) made by a method of drawing a mask on brass plates.

## 11. Discussion

In works of CERN the multichannel gas electron multipliers consisting of a plastic plate 50-2000 microns thick with metal or high-resistive thin (some microns) deposits from two parties are presented [2-4]. These GEMs provide the best spatial resolution and higher rate that wire chambers. However, an essential disadvantage of these GEM consists in their low reliability and stability. The matter is that at cathode dispersion by positive ions from proportional avalanches in GEM with metal or high-resistive electrodes there is a sedimentation of the sprayed carrying-out material on walls of holes takes place, which leads to subsequent leaks and breakdowns between electrodes.

In our works we have study the Penning mixtures: Ne + 30%$H_2$, Ar + 10%Xe, Ar + 10%$H_2$, 50%He + 50%Ne, Ar(Xe) + 20%$CF_4$, LAr(LXe) + 20%$CF_4$ [44,45]. It was shown for this mixtures that in proportional discharge besides the primary avalanches there is also the tail of secondary avalanches caused by metastable and nonmetastable Penning effects. For different mixtures of gases the duration of the signals caused by secondary avalanches are equal to $\sim$ n*10 microseconds (n = 1 - 10), the total charge of secondary avalanches is comparable or larger than in primary avalanches.

In Baksan laboratory we used a Penning mixture Xe + 0,8%$H_2$ at pressure 25 bar in proportional counter to search for double beta decay of $^{136}$Xe [49,50].

In this work we propose as another filling of the chambers for search of low-mas WIMP (<10 GeV/$c^2$) and solar axions on spin-dependent interaction with deuterium ($D_2$), $^3$He, $^{19}$F and $^{21}$Ne is used the mixtures: $^{21}$Ne + 10%$H_2$ [21], $D_2$ + 3ppmTMAE [25], $^3$He + 3%$CH_4$ [39] at pressure 10-17 bar. And in our experiment with liquid mixtures [16-18, 37, 44] is used the mixtures LAr + $CF_4$ ($^{19}$F) and LXe + $CF_4$ ($^{19}$F, $^{129}$Xe, $^{131}$Xe).

In work [46] is claimed that for targets with spin- nonzero nuclei it might be the spin-dependent interaction that determines the lower bound for the direct detection rate when the cross section of the scalar interaction, which is usually assumed to be the dominant part, drops below $10^{-12}$- $10^{-13}$ pb. In particular, from this work one can see that all fluorine-containing targets (LiF, $CF_4$, $C_2F_6$, and $CaF_2$, etc.) have almost the same sensitivity to both the SD and SI WIMP-nucleus interactions. Among all materials considered a detector with a $^{73}$Ge, $^{129}$Xe, or NaI target has better prospects to confirm or to reject the DAMA result [29] due to the largest values of the lower bounds for the total rate R(5, 50) > 0.06 -0.08 events/(kg day). If, for example, one ignores the SI WIMP interaction, then all materials have almost the same prospects to detect DM particles with the only exception of $CH_4$. In this context, in our experiments [15] with liquid $CH_4$ as another filling of the chamber for search of low-mas WIMP (<10 GeV/$c^2$) we propose to use the mixtures L$CH_4$ + 40ppmTMAE (section 7.2).

## 12. Summary & Outlook

In our works [9-14] GEMs with wire (WGEM) or metal electrodes (MGEM) and gas gap between metal electrodes without plastic were realized. An absence of a plastic insulation between electrodes of these GEMs excludes leakage currents and spark breakdowns between the electrodes.

In our works [16, 17, 21, 25] it was suggested to search low mass WIMPs and solar axions with help of chambers with GEMs and systems WGEM (MGEM) + pin-anodes. In work [18] we proposed a addition in liquid Xe (Ar) of photosensitive dopants and a comparison of scintillation (S1) and ionization signals (S2) for every event is suggested. In our works [15, 25, 37] it was suggested that the search for spin-dependent WIMP-nucleon interactions with help of detecting system GEM + pin-anodes can be performed.

In that respect we would like to add the next important comments:

1. As far as WIMPs with large masses (> 10 GeV) experimentally were not found so far [24, 27, 29, 31, 52], it is necessary to search the WIMP with small masses (≤ 10 GeV/$c^2$).

2. The data of the new DAMA/LIBRA–phase2 confirm a peculiar annual modulation of the single-hit scintillation events in the (2–6) keV energy range (WIMP mass < 10 GeV/$c^2$) satisfying all of the multiple requirements of the Dark Matter [29]. J.Va'vra have supposed [31] that this effect is explained by low mass WIMP (~1 GeV/$c^2$) scattering on protons in $H_2O$ molecules (H+).

3. In this context, in our experiments [15] with liquid $CH_4$ as filling of the chamber for search of low-mas WIMP we propose to use the mixture with $^1$H (L$CH_4$ + 40ppmTMAE). As well as, it is necessary to search of WIMP with small masses (≤ 10 GeV/$c^2$) in spin-dependent interactions between DM particles and gases with nonzero-spin nuclei ($D_2$, $^3$He, $^{19}$F, $^{21}$Ne, $^{129}$Xe, $^{131}$Xe) [46].

4. The chamber with system GEM + pin-anodes (section 5.5) and time projection chamber (section 8) with the mixture $D_2$ + 3ppmTMAE filling at a pressure 10-20 bar allow to search of spin-dependent interactions of solar axions and deuterium.

**References.**